\documentclass[12pt]{article}
\usepackage{amsthm,amsmath,latexsym,multibox,amssymb,amsfonts}
\usepackage{multibox}
\usepackage{graphicx,lscape,fancyhdr,array,stmaryrd}

\newcounter{Summa}
\newcounter{Csx}
\newcounter{Csy}
\newcounter{SizeX}

\newcommand{\Length}[1]{#10}
\newcommand{\SizeX}[2]{\setcounter{#1}{\Length{#2}}\addtocounter{#1}{12}}

\newcommand{\CellSize}{10}
\newcommand{\CellSizeE}{10.38}

\newcommand{\Cell} {15}
\newcommand{\Celld}{25}

\newcommand{\Inc}[2]{\setcounter{#1}{#2} \addtocounter{#1}{1} }

\newcommand{\CellFont} { \footnotesize }

\newcommand{\YoungBlock}[2] {
    \begin{picture}(\Length{#1},\Length{#2})(0,0)
    \Inc{Csx}{#1} \Inc{Csy}{#2}
    \multiput(0,0)(\CellSize,0){\value{Csx}}{\line(0,1){\Length{#2}}}
    \multiput(0,0)(0,\CellSize){\value{Csy}}{\line(1,0){\Length{#1}}}
    \end{picture}
  }

\newcommand{\YoungOneRowE}[2] {
    \SizeX{SizeX}{#1}
    \begin{picture}(\Length{#1},\CellSizeE)(10,#2)
    \put(0,0){ \YoungBlock{#1}{1} }
    \end{picture}
}

\newcommand{\YoungOneRow}[2] {
    \begin{picture}(\Length{#1},\Cell)(0,0)
    \multiframe(0,0)(10,0){1}(\Length{#1},\CellSize){{{\CellFont $#2$}}}
    \end{picture}
}

\newcommand{\YoungTwoRow}[4] {
    \begin{picture}(\Length{#1},\Celld)(0,0)
   \multiframe(0,\CellSizeE)(10,0){1}(\Length{#1},\CellSize){{{\CellFont$#2$}}}
   \multiframe(0,0)(10,0){1}(\Length{#3},\CellSize){{{\CellFont $#4$}}}
    \end{picture}
}

\newcommand{\YoungTwoRowE}[5] {
    \begin{picture}(\Length{#1},\Celld)(0,#5)
   \multiframe(0,\CellSizeE)(10,0){1}(\Length{#1},\CellSize){{{\CellFont$#2$}}}
    \multiframe(0,0)(10,0){1}(\Length{#3},\CellSize){{{\CellFont $#4$}}}
    \end{picture}
}

\newcommand{\YoungTwoRowEG}[3] {
    \begin{picture}(\Length{#1},30)(0,#3)
    \multiput(0,\CellSizeE)(10.3,0){#1}
    {\multiframe(0,\CellSize)(\CellSize,0){1}(\CellSize,\CellSize){} }
    \multiput(0,0)(10.3,0){#2}
    { \multiframe(0,\CellSize)(\CellSize,0){1}(\CellSize,\CellSize){} }
    \end{picture}
}

\newcommand{\YoungTwoRowBoldEG}[3] {
    \begin{picture}(\Length{#1},30)(0,#3)
    \linethickness{0.4mm}
    \multiput(0,\CellSizeE)(10.3,0){#1}
    { \multiframe(0,\CellSize)(\CellSize,0){1}(\CellSize,\CellSize){} }
    \multiput(0,0)(10.3,0){#2}
    { \multiframe(0,\CellSize)(\CellSize,0){1}(\CellSize,\CellSize){} }
    \end{picture}
}

\newcommand{\YoungThreeRowEG}[4] {
    \begin{picture}(\Length{#1},40)(0,#4)
    \multiput(0,20.6)(10.35,0){#1}
    { \multiframe(0,\CellSize)(\CellSize,0){1}(\CellSize,\CellSize){} }
    \multiput(0,\CellSizeE)(10.35,0){#2}
    { \multiframe(0,\CellSize)(\CellSize,0){1}(\CellSize,\CellSize){} }
    \multiput(0,0)(10.3,0){#3}
    { \multiframe(0,\CellSize)(\CellSize,0){1}(\CellSize,\CellSize){} }
    \end{picture}
}

\newcommand{\YoungA} [1] {\YoungOneRowE{1}{#1}}

\newcommand{\YoungB} [1] {\YoungOneRowE{2}{#1}}
\newcommand{\YoungC} [1] {\YoungOneRowE{3}{#1}}

\newcommand{\YoungAA}[1] {\YoungTwoRowEG{1}{1}{#1}}
\newcommand{\YoungBA}[1] {\YoungTwoRowEG{2}{1}{#1}}
\newcommand{\YoungBoldBA}[1] {\YoungTwoRowBoldEG{2}{1}{#1}}
\newcommand{\YoungCA}[1] {\YoungTwoRowEG{3}{1}{#1}}
\newcommand{\YoungDA}[1] {\YoungTwoRowEG{4}{1}{#1}}

\newcommand{\YoungBB}[1] {\YoungTwoRowEG{2}{2}{#1}}
\newcommand{\YoungCB}[1] {\YoungTwoRowEG{3}{2}{#1}}

\newcommand{\YoungAAA}[1] {\YoungThreeRowEG{1}{1}{1}{#1}}
\newcommand{\YoungBAA}[1] {\YoungThreeRowEG{2}{1}{1}{#1}}

\newcommand{\YoungCAA}[1] {\YoungThreeRowEG{3}{1}{1}{#1}}

\newcommand{\YoungOneRowB}[3] {
    \begin{picture}(\Length{#1},\Celld)(0,#3)
\multiframe(1.96,\CellSizeE)(10,0){1}(\Length{#1},\CellSize){{{\CellFont
$#2$}}}
    \YoungB{-0.2}
    \end{picture}
}

\newcommand{\YoungTwoRowBold}[4] {
    \begin{picture}(\Length{#1},\Celld)(0,0)
    \linethickness{0.4mm}
    \multiframe(0,11.1)(10,0){1}(\Length{#1},\CellSize){{{\CellFont $#2$}}}
    \multiframe(0,0)(10,0){1}(\Length{#3},\CellSize){{{\CellFont $#4$}}}
    \end{picture}
}
\newcommand{\YoungGeneralized }
{
    \begin{picture}(110,100)(0,0)
    \put(0,80) { \YoungBlock{9}{1} $n_1$}
    \put(0,70) { \YoungBlock{9}{1} $n_2$}
    \put(0,60){ \YoungBlock{8}{1} $n_3$}
    \put(4,30){ \line(0,1){30}}
    \put(10,47){$\vdots$}
    \put(10,33){$\vdots$}
    \put(0,20) { \YoungBlock{4}{1} $n_{p-2}$}
    \put(0,10) { \YoungBlock{4}{1} $n_{p-1}$}
    \put(0,0){ \YoungBlock{3}{1} $n_p$}
    \end{picture}
}

\newcommand{\beq}{\begin{equation}}
\newcommand{\eeq}{\end{equation}}
\newcommand{\bee}{\begin{eqnarray}}
\newcommand{\eee}{\end{eqnarray}}
\newcommand{\bml}{\begin{multline}}
\newcommand{\eml}{\end{multline}}
\newcommand{\bsplit}{\begin{split}}
\newcommand{\esplit}{\end{split}}
\newcommand{\ls}{\!\!\!\!\!\!}
\newcommand{\pl}{\partial}

\newcommand{\sigAp}{\sigma^1_+}
\newcommand{\sigBp}{\sigma^2_+}
\newcommand{\sigAm}{\sigma^1_-}
\newcommand{\sigBm}{\sigma^2_-}

\newcommand{\trace}[1]{{#1}'}
\newcommand{\traceless}[1]{ \mathbf \Pi \left[ #1\right]}
\newcommand{\young}[1]{\mathcal{P}\left[ \right]}

\newcommand{\AdS}{$\mbox{(A)dS}_d\,\,$}
\newcommand{\grLorenz}{${o(d-1,1)\,\,}$}
\newcommand{\grAdS}{${o(d-1,2)}\,\,$}

\newcommand{\n}[1]{\underline{#1}}
\renewcommand{\theequation}{\arabic{section}.\arabic{equation}}

\begin{document}

\begin{flushright}
\vspace{1mm}
FIAN/TD/20-05\\
\end{flushright}

\vspace{1cm}

\begin{center}
{\bf \Large Geometric Formulation for Partially Massless Fields}
\vspace{1cm}

\textsc{E.D. Skvortsov\footnote{skvortsov@lpi.ru} and M.A.
Vasiliev\footnote{vasiliev@lpi.ru}}

\vspace{.7cm}

{\em I.E.Tamm Department of Theoretical Physics, P.N.Lebedev
Physical
Institute,\\
Leninsky prospect 53, 119991, Moscow, Russia}

\vspace{3mm}

\end{center}

\vspace{2cm}

\begin{abstract}
The manifestly gauge invariant formulation for free symmetric
higher-spin partially massless fields in $(A)dS_d$ is given in
terms of gauge connections and linearized curvatures  that take
values in the irreducible representations of  $(o(d-1,2))\ o(d,1)$
described by two-row Young tableaux, in which the lengths of
 the first  and second row are, respectively,
 associated with the spin and depth of partial masslessness.

\end{abstract}

\vspace{0.2cm}

\begin{center}
\vspace{0.1cm} {Keywords: Higher-spin
fields gauge theory; partially-massless fields; (anti)-de Sitter
space;} \\ ArXiv.org:hep-th/0601095
\end{center}

\setcounter{footnote}{0}

\newpage

\tableofcontents \vspace{2cm}

\newpage

\section{Introduction}
\setcounter{equation}{0}
 An interesting property of massive
higher-spin (HS) fields in spaces with nonzero curvature
 like de Sitter ($dS$) and anti-de Sitter ($AdS$) is that
for certain values of the parameter of  mass, the HS field
equations acquire gauge invariance. This phenomenon was originally
discovered  by Deser and Nepomechie in \cite{FirstPMWorks} and
investigated further in \cite{Higuchi}-\cite{Dolan:2001ih}. The
special values of the mass parameter are scaled in terms of the
curvature of the background space. The first special value of the
mass corresponds to the usual massless fields with the richest
gauge symmetries. The other special values correspond to different
fields called partially massless fields. A gauge transformation
law of a partially massless field has less parameters but more
derivatives compared to the massless fields. A number of degrees
of freedom carried by a partially massless field is intermediate
between the massless case (two for $s>0$ in $4d$) and the  massive
case ($2s+1$ in $4d$). In the flat space limit all special values
of the mass shrink to zero so that one is left with the usual
massless fields, i.e., partially massless theories do not exist in
Minkowski space.

In $AdS_d$ space-time, the phenomenon of partial masslessness has
clear interpretation in terms of representations of the $AdS_d$
algebra $o(d-1,2)$. Namely, a space of single-particle states of a
given relativistic field with the energy bounded from below forms
a lowest weight
 $o(d-1,2)$-module, $|E_0,s \rangle$. Its lowest weight is defined in
terms of the lowest energy $E_0$ and  spin $s$ associated with the
weights of the maximal compact subalgebra $o(2)\oplus o(d-1)
\subset o(d-1,2)$. For quantum-mechanically consistent fields, the
modules $|E_0,s \rangle$  correspond to unitary representations of
$o(d-1,2)$ that requires in particular $E_0 \geq E_0 (s)$ where
$E_0 (s)$ is some function of the spin $s$ found for the case of
$d=4$ in \cite{Evans} and for the general case in
\cite{Metsaev:1997nj} (see also \cite{Angelopoulos:1997ij,
Brink:2000ag, Ferrara:2000nu, Vasiliev:2004cm, Dolan:2005wy}). At
$E_0 = E_0 (s)$ null states appear that form a submodule to be
factored out. These signal a gauge symmetry of the system. The
corresponding fields are the usual massless fields. Modules with
lower energies $E< E_0(s)$ correspond to nonunitary (ghost)
massive fields. At certain values $E_i (s)$ of $E$ the
corresponding nonunitary module may contain a submodule that again
signals a gauge symmetry in the field-theoretical description.
Such modules correspond to the partially massless fields.
Unfortunately, the lack of unitarity \cite{Dolan:2001ih} makes
their physical applications problematic in the $AdS$ case.

In the $dS_d$ case with the symmetry algebra $o(d,1)$ the analysis
is different because it admits no unitary lowest weight (i.e.,
bounded energy) representations at all. Relaxing the concept of
bounded energy, it is possible to have unitary partially massless
fields in that case \cite{Higuchi, DeserWaldron, Dolan:2001ih}.

In this paper we address the question what is a frame-like
formulation of the symmetric  partially massless fields, that
generalizes the frame-like formulation of ordinary symmetric
massless fields elaborated in \cite{Vasiliev:1980as,
Vasiliev:1986td, Lopatin:1987hz,Vasiliev:2001wa}.
 Such a reformulation operates in terms of a
gauge field (connection), that takes values in some module of the
$(A)dS_d$  algebra and makes gauge symmetries of the system
manifest allowing to write down manifestly gauge invariant
geometric actions. As shown in \cite{Vasiliev:2001wa}, a usual
spin--$s$ massless field in $(A)dS_d$ is described by a 1-form
connection, which takes values in the finite-dimensional
$o(d-1,2)$-module spanned by traceless tensors of  $o(d-1,2)$ that
have a symmetry of a length $s-1$ two-row rectangular Young
tableau $\YoungTwoRowBold{4}{s-1}{4}{s-1}\ .$ The main result of
this paper is that a spin--$s$ partially massless symmetric gauge
fields is analogously described by a 1-form connection, which is a
$o(d-1,2)(o(d,1))$-tensor that has the symmetry properties of a
two-row Young tableaux with the first row of length $s-1$ and the
second row of an arbitrary length $t= 0,1\ldots s-1$,
$\YoungTwoRowBold{5}{s-1}{3}{t}.$

Since the gauge fields of the frame-like formalism are interpreted
as connections of a non-Abelian HS algebra of a full HS system,
the results of such a reformulation may be useful  to rule out HS
algebras incompatible with unitarity by requiring that physically
relevant algebras considered as $o(d-1,2)$-modules under its
adjoint action should not contain the modules  associated with
(nonunitary) partially massless fields. (Note that, as shown in
\cite{Vasiliev:2001wa}, \cite{Alkalaev:2003qv}, gauge fields
associated with true massless fields in $AdS_d$ carry $o(d-1,2)$
indices of various $o(d-1,2)$-modules described by Young tableaux,
that have first two rows of equal lengths.) Although in this paper
we consider integer spins, our approach admits a straightforward
extension to  fermions.

The layout of the rest of the paper is  as follows.  In Section 2
we recall the metric-like formulation of massive, massless and
partially massless fields. The frame-like formalism for massless
HS fields is recalled  in Section 3. The main results of our
approach are presented in Section 4 and then illustrated in
Section 5 by the spin-2 and spin-3 examples. Possible further
developments are discussed  in Section 6. Notation, Young tableau
convention and explicit expressions for some coefficients are
collected in appendices.

\section{Metric Formalism for Free Higher-Spin Fields  }
\setcounter{equation}{0} \label{Overview} In this section we
summarize known results on the description of massive, massless
and partially massless HS fields, which are relevant to our
analysis.

\subsection{Massive higher-spin fields}
A  massive spin-$s$ totally  symmetric field in the flat Minkowski
space can be described by a rank-$s$ symmetric tensor field
$\phi^{\{s\}}_{\underline{m}_1\ldots\underline{m}_s}(x) $ that
satisfies the equations \cite{Dirac:1936}\footnote{Our notations
and conventions are explained  in Appendix~A.} \beq\begin{split}
\label{MassiveField} &(\square+m^2)\phi^{\{s\}}=0,
\\ &\pl\cdot\phi^{\{s\}}=\mbox{divergence}(\phi^{\{s\}})=0,\quad s \geq1,  \\
&\trace{\phi^{\{s\}}}=\mbox{trace}(\phi^{\{s\}})=0, \quad
s\geq2\,,\end{split}\, \eeq which form a complete set of local
Lorentz-invariant conditions that can be imposed on
$\phi^{\{s\}}$. The first equation of (\ref{MassiveField}) is the
mass-shell condition. The third condition, that the field
$\phi^{\{s\}}$ is traceless, is an
 algebraic constraint. As pointed out by Fierz and
Pauli in the pioneering paper \cite{Fierz:1939ix},  the second
condition, which is a differential equation, can be derived
 for $s>1$ from a Lagrangian without introducing new degrees of freedom by
adding  auxiliary fields that are expressed in terms of
(derivatives of) the dynamical fields by virtue of equations of
motion. For totally symmetric massive fields of integer spins, the
lagrangian formulation  with a minimal set of auxiliary fields was
worked out by Singh and Hagen in \cite{Singh:1974qz}. For a spin
$s$ they introduced the set of auxiliary fields  that
 consists of symmetric traceless tensors of ranks
$s-2, s-3,\ldots 0$.

For example, to describe a spin-2 field, one auxiliary scalar
field is needed. It can be identified  with the trace of a
traceful rank-2 field $\phi_{\underline{m}\underline{n}}$
satisfying the Fierz-Pauli equation \beq \label{SpinTwo}
G_2=(\Box+m^2)(\phi_{\underline{m}\underline{n}}-\eta_{\underline{m}
\underline{n}}\trace{\phi})+
\pl_{\underline{m}}\pl_{\underline{n}}\trace{\phi}+\eta_{\underline{m}
\underline{n}}\pl^{\underline{l}}\pl^{\underline{p}}\phi_{\underline{l}
\underline{p}}-
2\pl_{(\underline{m}}\pl^{\underline{k}}\phi_{\underline{k}
\underline{n})}= 0.\eeq

One obtains \beq\label{SpinTwo1}\pl\cdot
G_2=m^2(\pl\cdot\phi-\pl\trace{\phi})\eeq and \beq
\label{SpinTwo2} \pl^2\cdot
G_2-\frac{m^2}{d-2}\trace{G_2}=m^4\frac{d-1}{d-2}\trace{\phi}.\eeq
{}From equation  (\ref{SpinTwo2}) it follows that, if $m^2\neq 0$,
$\phi$ is traceless (i.e., the auxiliary field is zero on-shell).
Then from (\ref{SpinTwo1}) it follows that $\pl\cdot\phi =0$.

If $m^2 =0$, equation (\ref{SpinTwo1}) expresses  Bianchi
identities which manifest the spin-2 gauge symmetry. Let us note
that it is important that the higher-derivative part of the
massive equations is the same as for massless fields just to make
it possible to express the auxiliary fields in terms of the
dynamical fields by virtue of algebraic constraints, that would
not be possible for a different choice of the coefficients
 in front of the second
derivative terms in (\ref{SpinTwo}). This means that, for
 a different choice of the coefficients, the corresponding equations
describe a dynamical system with the propagating trace field, that
possesses more degrees of freedom. A similar argument will  play
crucial role in the subsequent analysis of a proper action for a
partially massless field.

Note that equivalent dynamical systems
 can be formulated with the aid of different sets of
auxiliary fields. Particularly useful approaches to massive fields
with different sets of auxiliary fields include the gauge
invariant formulation of \cite{Zinoviev}, \cite{Bianchi:2006gk}
and the BRST approach of \cite{Buchbinder:2005ua,
Buchbinder:2005cf}.

\subsection{Massless higher-spin fields}

The free action for a symmetric massless field of an arbitrary
integer spin in the flat Minkowski space was obtained by Fronsdal
\cite{Fronsdal:1978rb}. A symmetric spin-$s$ massless field is
described by a totally symmetric double traceless tensor field
$\phi^{\{s\}}\equiv\phi_{\underline{m}_1 \ldots
\underline{m}_s}(x)$, $\trace{\trace{\phi^{\{s\}}}}=0$. The
equations of motion \beq
\mathcal{F}^s=\square\phi^{\{s\}}-\pl\pl\cdot\phi^{\{s\}}+\pl^2\trace{
{\phi^{\{s\}}}}=0\eeq are invariant under gauge transformations
\beq\label{FronsdalGauge} \delta\phi^{\{s\}}=\pl\xi^{\{s-1\}}\eeq
with a gauge parameter $\xi^{\{s-1\}}(x)$ being a symmetric
rank-$(s-1)$
 traceless tensor field. The gauge invariant Lagrangian is
 \beq \mathcal{L}^s=\frac12
\phi^{\{s\}}\cdot\mathcal{F}^s-\frac18
s(s-1){\trace{\phi^{\{s\}}}}\cdot \trace{\mathcal{F}^s}\,.
\label{Fronsdal}\eeq Note that alternative metric-like
formulations of symmetric massless HS fields were also developed
in \cite{Buchbinder:2005cf, Buchbinder:2004gp,
MasslessHSAlternative}.

\subsection{Partially massless higher-spin fields}\label{PMreview}

 The \AdS space is described by  a metric
$g_{\n{m}\n{n}}$ satisfying the equation \beq
R_{\n{m}\n{n},\n{p}\n{q}}=-\frac{2\Lambda}{(d-1)(d-2)}(g_{\n{p}\n{m}}
g_{\n{n}\n{q}}-g_{\n{p}\n{n}}g_{\n{m}\n{q}})\,, \eeq where
$R_{\n{m}\n{n},\n{p}\n{q}}$ is the Riemann tensor and the
normalization of the cosmological constant is chosen as in
\cite{DeserWaldron, Zinoviev}. $\Lambda$ is (negative)positive for
(anti-)de Sitter space.  Later on we use the notation
$\lambda^2=-\frac{2\Lambda}{(d-1)(d-2)}$, where $|\lambda^{-1}|$ is
the  \AdS radius.

\subsubsection{The approach of Deser and Waldron}
\label{The approach of Deser and Waldron}

In this subsection, we consider the case of $d=4$  following to
\cite{DeserWaldron} (modulo the adjustment of the coefficients to the
mostly minus signature conventions used in this paper). To illustrate
the effect of partial masslessness let us start with the simplest
example of spin two.

The deformation of the massive field equation
 (\ref{SpinTwo}) to \AdS is achieved by the
addition of some mass-like terms proportional to $\Lambda$ along
with the replacement of the flat derivatives by the covariant ones
$D$ associated with $g_{\n{m}\n{n}}$.

The \AdS\ deformation of the equation (\ref{SpinTwo}) is \beq G_2
= (\Delta^{(2)}+m^2-\Lambda)(\phi-g\trace{\phi})-\Lambda\phi+DD
\trace{\phi}-DD\cdot\phi+gD\cdot D\cdot\phi=0,\eeq where  we use
the notation of Deser and Waldron \cite{DeserWaldron} \beq
\Delta^{(2)}\phi=\square\phi+\frac{8\Lambda}3(\phi-\frac14 g
\trace{\phi}). \eeq The equations (\ref{SpinTwo1})
 and (\ref{SpinTwo2}) modify to
  \beq
\label{SpinTwoBianchi}D\cdot G_2=m^2(D\cdot\phi-D\trace{\phi})\,,
\eeq \beq D^2\cdot G_2-\frac12
m^2\trace{G_2}=\frac12m^2(3m^2-2\Lambda)\trace{\phi}.
\label{SpinTwoTrace} \eeq

One observes that, for the critical value  $m^2=\frac23\Lambda$,
$\phi^\prime$ cannot be algebraically expressed in terms of the
other field and the equation (\ref{SpinTwoTrace}) becomes  Bianchi
identity signaling the appearance of a gauge invariance which is
\beq \label{PMSpinTwoGaugeLaw}
\delta\phi_{\n{m}\n{n}}=(D_{(\n{m}}D_{\n{n})}-\frac{\Lambda}{3}
g_{\n{m}\n{n}})\xi \eeq with a scalar gauge parameter $\xi$. This
invariance kills the helicity zero degree of freedom leaving spin
polarizations $\pm1,\pm2$. Note that although $\phi^\prime$ cannot
be expressed in terms of the other field, this does not imply the
appearance of extra degrees of freedom because it is pure gauge.

At $m=0$, (\ref{SpinTwoBianchi}) is the Bianchi identity of the
usual spin-2 massless theory in \AdS.

A spin-3 massive field requires a scalar auxiliary field $\chi$
and an auxiliary vector, which combines with a rank-3 traceless
field into a rank-3 traceful tensor $\phi_{\underline{m}
\underline{n} \underline{k}}$. The equations of motion of the
massive spin-3 field $\phi_{\underline{m} \underline{n}
\underline{k}}$ are \cite{DeserWaldron}
 \beq
\begin{split}
G_3=(&\Delta^{(3)}+m^2-\frac{16\Lambda}{3})\phi-DD\cdot\phi+DD\trace{\phi}-
\\ &-g\Big((\Delta^{(1)}+m^2-\frac{11\Lambda}{3})\trace{\phi}-D\cdot
D\cdot\phi+\frac12
DD\cdot\trace{\phi}\Big)-\frac{m}4gD\chi=0,\end{split}\eeq \beq
G_0=\frac32(\Delta^{(0)}+4m^2-8\Lambda)\chi-mD\cdot \trace{\phi}=0
\eeq with \beq
\Delta^3\phi=\square\phi+5\Lambda\phi-\frac{2\Lambda}3 g
\trace{\phi}\,,\qquad
\Delta^{(1)}\phi^\prime=\square\phi^\prime+\Lambda\phi^\prime
\,,\qquad \Delta^{(0)}\chi=\square\chi\,. \eeq The following
identities hold
 \beq
\label{SpinThree2}\traceless{D\cdot
G_3}=\frac12m\traceless{(-DD\chi-m D\trace{\phi}+ m D\cdot
\phi)},\eeq \beq \label{SpinThree1} D^2\cdot G_3 +\frac{m}4 D G_0
- \frac{m^2}4\trace{G_3} = \frac58m(3m^2-4\Lambda)(D\chi-\frac23m
\trace{\phi}),\eeq \beq \label{SpinThree0} D^3\cdot G_3 -
\frac5{12}m(3m^2-4\Lambda)G_0 =-\frac52
m(3m^2-4\Lambda)(m^2-2\Lambda)\chi\,, \eeq where $\traceless{...}$
is the projector to the traceless part.

  At $m=0$, (\ref{SpinThree2})
becomes the Bianchi identity associated with the gauge invariance
with a rank-2 traceless parameter $\xi^{\{2\}}$ \beq
\delta\phi=D\xi^{\{2\}},\qquad \ \ \delta \chi=0. \eeq In this
case, the system decomposes into two independent subsystems:
massless spin three with two spin projections $\pm3$ is described
by the Fronsdal field $\phi$ and  massive spin zero is described
by $\chi$.

At $m^2=\frac43\Lambda$, (\ref{SpinThree1}) becomes the Bianchi
identity associated with the gauge invariance with the rank-1
gauge parameter $\xi^{\{1\}}$, which has the form

\beq\label{SpinThreeDepthOne} \delta\phi=(D^2-\frac{\Lambda}3g
)\xi^{\{1\}},\qquad \ \ \delta
\chi=-\frac23\sqrt{\frac{\Lambda}3}D\cdot\xi^{\{1\}}. \eeq This
model describes a field with spin projections $\pm2,\pm3$.

At $m^2=2\Lambda$, (\ref{SpinThree0}) becomes the Bianchi identity
that manifests the gauge invariance with the scalar parameter
$\xi^{\{0\}}$
 \beq
\delta\phi=(D^3-\frac{\Lambda}2g D)\xi^{\{0\}},\qquad \ \ \delta
\chi=-\frac23\sqrt{\frac{\Lambda}3}(D\cdot
D-\frac{10\Lambda}3)\xi^{\{0\}}, \eeq leaving out the helicity
zero degree of freedom. At  non-critical $m^2$ the scalar  $\chi$
and vector $\trace{\phi}$ are auxiliary. They are zero on shell
  and  $D\cdot\phi=0$ as a consequence of the constraints
(\ref{SpinThree2}), (\ref{SpinThree1}) and (\ref{SpinThree0}). A
generic non-critical system describes  a massive spin-3 field with
spin projections $(0,\pm 1,\pm 2,\pm 3)$ (may be ghost and/or
tachionic for negative $m^2$).

For general spin $s$, the full set of fields contains a rank-$s$
double traceless field along with the traceless fields of ranks
from $0$ to $s-3$. As argued in \cite{DeserWaldron}, there exist
$s$ critical masses ${m_t}^2$
 ($t=0,1,..., s-1$)
and the following properties are true for $m^2$= ${m_t}^2$:
\begin{itemize}
     \item
The equations of motion are invariant under the gauge
transformation
     \beq
\label{gtr} \delta \phi^{\{s\}}=\underbrace{D\ldots
     D}_{s-t}\xi^{\{t\}}+\ldots
     \eeq
with $s-t$ derivatives and  a rank-$t$ traceless gauge parameter
$\xi^{\{t\}}$.
     \item A  rank-$t$ traceless combination of the equations of motion
     becomes a Bianchi identity.
     \item The physical spin projections are
     $\pm(t+1), \ldots, \pm s$ plus decoupled lower spin massive fields.
\end{itemize}

\subsubsection{The approach of Zinoviev} \label{ZinovievApproach}

A useful approach to the description of massless, massive and
partially massless  fields in $(A)dS_d$ was proposed by Zinoviev
in \cite{Zinoviev}. The Zinoviev's lagrangian is the sum of the
covariantized Fronsdal's lagrangians $\mathcal{L}^s_{{cov}}$
(\ref{Fronsdal}) (with flat derivatives replaced by the covariant
ones) for the set of spin-$k$ massless fields $\phi^{\{k\}}$ with
$k=0\ldots s$ plus some lower-derivative terms, i.e.,
 \beq
\mathcal{L}^s=\sum_{k=0}^{k=s}
\mathcal{L}^k_{{cov}}+\triangle\mathcal{L}^s
\label{Lagrang}\,,\eeq where
 \beq
\begin{split} \label{LowDerPart}
\triangle\mathcal{L}^s =\sum_{k=0}^{k=s} \Big (
a_k&\phi^{\{k-1\}}({D}\cdot\phi^{\{k\}})+
b_k\trace{\phi^{\{k\}}}({D\cdot}\phi^{\{k-1\}})+
c_k({D}\cdot\trace{\phi^{\{k\}}})\trace{\phi^{\{k-1\}}}+
\\&+d_k({\phi^{\{k\}}})^2+e_k({\trace{\phi^{\{k\}}}})^2+f_k\trace{\phi^{\{k\}}}
\phi^{\{k-2\}}\Big ).
\end{split} \eeq
The  coefficients in $\triangle\mathcal{L}^s$ are fixed by the
condition that $\mathcal{L}^s$ is invariant under some gauge
transformations of the form \beq \label{GaugeLaw}
\delta\phi^{\{k\}}=\frac1k
D\xi^{\{k-1\}}+\alpha_k\xi^{\{k\}}+\beta_k g\xi^{\{k-2\}} \eeq
with rank-$k$ traceless gauge parameters $\xi^{\{k\}}$
($k=0,1,..., s-1$). The gauge invariance condition expresses all
the coefficients in terms of one free mass parameter $m^2$.

For generic $m^2$, the lagrangian (\ref{Lagrang}) describes a
spin-$s$  massive field. In that case all  gauge symmetries are
Stueckelberg (i.e., algebraic for some of the fields). It is easy
to see that the gauge fixing of all gauge symmetries reproduces
the set of fields of Singh and Hagen \cite{Singh:1974qz}, i.e.,
the Singh-Hagen theory deformed to $(A)dS_d$ is a gauge-fixed
version of the theory of Zinoviev. In this approach it is still
unclear, however, what is the fundamental principle that underlies
the gauge symmetry of the model, thus fixing the coefficients in
the action (\ref{Lagrang}).

At the critical values of the mass \cite{Zinoviev}
  \beq
{m^2}_t=\frac{2\Lambda}{(d-1)(d-2)}(s-t-1)(d+s+t-4),\ \ t=0,1,...,
s-1\eeq some of the coefficients in the gauge transformation law
(\ref{GaugeLaw}) and in $\triangle\mathcal{L}^s$ vanish so that
the lagrangian (\ref{Lagrang}) splits into two independent parts.
One contains the fields $\phi^{\{k\}}$ with $k=t+1\ldots s$
 and another one contains those with $k=0 \ldots t$. The
former lagrangian describes a partially massless theory, including
the usual massless theory as a particular case of $t=s-1$, while
the latter describes a spin-$t$ massive field.  In the flat limit
$\Lambda \rightarrow 0$ all critical masses tend to zero and all
lower derivative terms in the lagrangian vanish so that
$\mathcal{L}^s$ reduces to the  sum of the flat space Fronsdal's
lagrangians of massless fields $\phi^{0},\ldots,\phi^s$ of all
spins from 0 to $s$.

\section{Massless Higher-Spin Fields in the Frame-Like Formalism}
\setcounter{equation}{0}
\subsection{Lorentz covariant approach}
The Minkowski or \AdS\ geometry can be described by the frame
field (vielbein) $e^a =dx^{\underline{m}} e_{\underline{m}}{}^a$
and  Lorentz spin-connection 1-form $\omega^{a,b} =
dx^{\underline{m} } \omega_{\underline{m}}{}^{a,b}$
($\omega^{a,b}$=-$\omega^{b,a}$), which obey the equations
\beq\label{ZeroTorsionA} T^a\equiv de^a+{\omega^{a,}}_b\wedge
e^b=0,\eeq \beq \label{ZeroCurvatureA} R^{a,b}\equiv
d\omega^{a,b}+{\omega^{a,}}_c\wedge\omega^{c,b}-{\lambda}^2e^a\wedge
e^b=0,\eeq  where $d=dx^{\underline{m}} \frac{\partial}{\partial
x^{\underline{m}}}$ is the de Rham operator, $-\lambda^2$ is the
cosmological constant and the exterior product symbol $\wedge$
will be systematically omitted in what follows.
 The equation (\ref{ZeroTorsionA}) is
the zero-torsion constraint that expresses the Lorentz connection
${\omega_{\underline{m}}}^{a,b}$ via vielbein
${e_{\underline{m}}}^a$, provided that the vielbein is
nondegenerate, which is required.
 The equation (\ref{ZeroCurvatureA}) then describes the $AdS_d$
space with the symmetry algebra $o(d-1,2)$ if $\lambda^2>0$,
$dS_d$ space with the symmetry algebra $o(d,1)$ if $\lambda^2<0$
and flat Minkowski space if $\lambda^2=0$. The latter corresponds
to the contraction of either $o(d-1,2)$ or $o(d,1)$ to
$iso(d-1,1)$ and can be described  by
$\omega_{\underline{m}}{}^{a,b}=0$ and
${e_{\underline{m}}}^a={\delta_{\underline{m}}}^a$ in Cartesian
coordinates. Dynamical gravity is described by the vielbein and
Lorentz connection which satisfy the zero-torsion condition
(\ref{ZeroTorsionA}) and   the equations (\ref{ZeroCurvatureA})
relaxed to the Einstein equations.

The Lorentz covariant derivative $D$ acts as usual \footnote{As
explained in more detail in Appendix A, upper or lower indices
denoted by the same letter are automatically symmetrized. }
\beq\begin{split}
DC^{a(n),b(m),\ldots}&=dC^{a(n),b(m),\ldots}+n{\omega^{a,}}_c
C^{ca(n-1),b(m),\ldots}+\\&+m{\omega^{b,}}_c
C^{a(n),cb(m-1),\ldots}+\ldots,\end{split}\eeq where
$C^{a(n),b(m),...,}(x)$ is an arbitrary Lorentz tensor field. As a
consequence of (\ref{ZeroCurvatureA}), one obtains \beq
D^2C^{a(n),b(m),\ldots} = \lambda^2 (n e^a e_c
C^{ca(n-1),b(m),\ldots}+ m e^b e_c C^{a(n),cb(m-1),\ldots} +\ldots
)\,. \eeq

Let us now recall how this approach extends to massless HS fields
in flat and $(A)dS$ spaces. The generalization to  HS massless
fields in the Minkowski space \cite{Vasiliev:1980as} starts with
the gauge fields ${e_{\underline{m}}}^{a(s-1)}$ and
${\omega_{\underline{m}}}^{a(s-1),b}$ with the tangent Lorentz
indices possessing the symmetry properties of traceless Young
tableaux of the form \YoungOneRow{5}{s-1} and
 \YoungTwoRow{5}{s-1}{1}{}, respectively\footnote{I.e. the 1-form $\omega^{a(s-1),b}$ is
symmetric in the indices $a$, traceless in the indices $a$ and
$b$, and the symmetrization over all indices gives zero:
$\omega^{a(s-1),a}\equiv0$. See also Appendix B.}. In Cartesian
coordinates, the gauge transformations are \beq
\begin{split} \label{HSGaugeLaw}&\delta
{e_{\underline{m}}}^{a(s-1)}=\pl_{\underline{m}}
\xi^{a(s-1)}+\xi^{a(s-1),}_{\phantom{a(s-1),}\underline{m}}\\
&\delta {\omega_{\underline{m}}}^{a(s-1),
b}=\pl_{\underline{m}}\xi^{a(s-1), b}+\xi^{a(s-1),b}{}_{\n{m}},
\end{split} \eeq
where the traceless gauge parameters $\xi^{a(s-1)}$,
$\xi^{a(s-1),b}$ and $\xi^{a(s-1),bb}$ have the symmetry
properties of  Young tableaux \YoungOneRow{4}{s-1},
\YoungTwoRow{4}{s-1}{1}{} and \YoungOneRowB{4}{s-1}{},
respectively.

In these terms the gauge invariant action can be written in the
following simple form \beq
\label{fract} S= \int_{M^d}\,
dx^{a_1} \ldots dx^{{a_{d-3}}}\,\varepsilon_{a_1 \ldots
a_{d-3}pqr}
(d\,e^{n_1 \ldots n_{s-2} p} +\frac{1}{2}\, dx^m \,\omega^{n_1
\ldots n_{s-2} p,}{}_{ m})\, \omega_{n_1 \ldots n_{s-2}}{}^{q,\,
r}.
\eeq It is easy to check its gauge invariance, although it is not
manifest. \label{AdSHS}

Further analysis \cite{Vasiliev:1986td, Lopatin:1987hz} shows
that, both in Minkowski and in \AdS\ cases, HS fields are
conveniently described by a set of 1-forms
${\omega_{\underline{m}}}^{a(s-1),b(t)}$, $t=0,1,...,s-1$,
possessing the symmetry properties of the two-row traceless Young
tableaux\footnote{I.e. $\omega^{a(s-1),b(t)}$ is symmetric in
indices $a$ and $b$ separately, traceless in all indices and
$\omega^{a(s-1),ab(t-1)}\equiv 0$. } \YoungTwoRow{4}{s-1}{3}{t}.
The higher connections ${\omega_{\underline{m}}}^{a(s-1),b(t)}$
with $t\geq 2$ called {\it extra fields} are gauge fields for
higher gauge symmetries the simplest of which is that with the
gauge parameter $\xi^{a(s-1),bb}$ in the gauge transformations
(\ref{HSGaugeLaw}). This leads to the gauge field
$\omega^{a(s-1),bb}$ which in its gauge transformation contains a
new gauge parameter thus leading to the next extra field  and so
on.

The gauge transformation law for
${e_{\underline{m}}}^{a(s-1)}\equiv
{\omega_{\underline{m}}}^{a(s-1)}$ is
 \beq \delta
{\omega_{\underline{m}}}^{a(s-1)}=D_{\underline{m}}\xi^{a(s-1)}+
e_{\underline{m}c} \xi^{a(s-1),c} \label{GaugeFronsdal},\eeq where
$D_{\underline{m}}$ is the Lorentz covariant derivative defined
with respect to the background spin connection $\omega^{a,b}$ and
vielbein $e^a$ that describe either Minkowski or (anti-)de Sitter
geometry. Raising a form index by the vielbein $\omega^{a(s-1)|b}=
{\omega_{\underline{m}}}^{a(s-1)}e^{\underline{m}|b}$ one observes
that this field decomposes into the following irreps of \grLorenz\
\beq \omega^{a(s-1)|b}=\YoungOneRow{5}{s}\
\oplus\YoungOneRow{4}{s-2}\ \oplus\YoungTwoRowE{4}{s-1}{1}{}{10}\
.\eeq The first two components  combine into the Fronsdal field: a
rank-$s$ double traceless tensor decomposes into traceless tensors
of ranks $s$ and $(s-2)$. The last component is Stueckelberg. It
is gauged away by the traceless gauge parameter $\xi^{a(s-1),c}$
that has the same symmetry type. As a result only the  gauge
parameter $\xi^{a(s-1)}$ is left and (\ref{GaugeFronsdal}) gives
rise to the Fronsdal gauge transformation law
(\ref{FronsdalGauge}).

With some rescaling of fields by $\lambda$ the gauge
transformations of the other connections have the form \beq
\delta{\omega_{\underline{m}}}^{a(s-1),b(t)}=D_{\underline{m}}
\xi^{a(s-1),b(t)}+\lambda e_{\underline{m}
c}\xi^{a(s-1),b(t)c}+\lambda a_{s,t}(d)
\mathcal{P}({\xi^{a(s-1),b(t-1)}{e_{\underline{m}}}^b)}\,, \eeq
where $a_{s,t}(d)$ are some coefficients and the projector
$\mathcal{P}$ restores the tracelessness and the proper Young
symmetry. All the parameters $\xi^{a(s-1),b(t)}$ with $t>0$ are
Stueckelberg, eliminating some components of the connections.

The field-strengths, which have analogous form\footnote{A label in
lower braces indicates the rank of a form; see Appendix A. } \beq
R^{a(s-1),b(t)}_{\{2\}} =D\omega^{a(s-1),b(t)}_{\{1\}}+\lambda
e_{c}\omega^{a(s-1),b(t)c}_{\{1\}}+ \lambda a_{s,t}(d)
 \mathcal{P}( {\omega^{a(s-1),b(t-1)}_{\{1\}}e^b})\,,
\eeq are demanded to be gauge invariant, \beq
\label{HSCurvatureInvariance}\delta R^{a(s-1),b(t)}_{\{2\}} =0\,,
\eeq which condition fixes the coefficients $a_{s,t}(d)$.

The frame-like HS field ${\omega_{\underline{m}}}^{a(s-1)}$ is
called dynamical. The rest of the fields
${\omega_{\underline{m}}}^{a(s-1),b(t)}$, $t=1,2,...,(s-1)$ are
expressed in terms of derivatives of the dynamical ones by virtue
of certain constraints \cite{Lopatin:1987hz} which set to zero
some components of the HS curvatures $R^{a(s-1),b(t)}_{\{2\}}$. A
length $t$  of the second row of the Young tableau associated with
a particular  connection equals to the highest order of
derivatives of the dynamical field in the resulting expressions
for the higher connections \beq
\omega^{a(s-1),b(t)}\sim{\left(\frac{\pl}{\lambda\pl
x}\right)}^t\omega^{a(s-1)} +\ldots\ . \eeq For example, in the
spin-2 case, the field $\omega^{a,b}$ is the Lorentz connection.
It is expressed in terms of the first derivatives of the dynamical
field, the vielbein $e^a$. The field $\omega^{a(s-1),b}$ is
analogous to the Lorentz connection and is called auxiliary.  The
condition that the free action contains at most two space-time
derivatives of the dynamical field is equivalent to the {\it extra
field decoupling condition} that requires the variation of the
free action with respect to the {\it extra fields} to vanish
identically
 \beq
\label{HS_EFDC_Lorentz} \frac{\delta S}{\delta
\omega^{a(s-1),b(t)}}\equiv0\,,\qquad t\geq2. \eeq

A general $P$-even, gauge invariant action in \AdS\ has the form
\cite{Vasiliev:1986td, Lopatin:1987hz} \bee S= \frac{1}{2\lambda}
\sum_{k=0}^{k=s-2} a_{q} \int_{M^d}
  \epsilon_{c_1 \ldots c_d} e^{c_5} \ldots e^{c_d}
 {R_{\{2\}}}^{c_1 a(s-q-2),c_2 b(s-2)}
{R_{\{2\}}}^{c_3\phantom{a(s-q-2)},c_4
}_{\phantom{c_3}a(s-q-2)\phantom{,c_4}b(s-2)}\,. \eee The extra
field decoupling condition fixes the coefficients $a_q$ modulo a
spin-dependent normalization factor. The resulting action is
manifestly gauge invariant by (\ref{HSCurvatureInvariance}),
containing just two derivatives of the dynamical field upon
excluding the auxiliary field by virtue of its equation of motion.
As such, it is equivalent to the Fronsdal action in $AdS_d$. (In
fact, the free action for massless fields in $AdS_d$ for any $d$ was
originally obtained in this form in \cite{Lopatin:1987hz}).

\subsection{\AdS covariant approach}
 The equations (\ref{ZeroTorsionA}) and (\ref{ZeroCurvatureA})
are equivalent to the zero-curvature condition  for the 1-form
connection ${W_0}^{A,B}$ of $o(d-1,2)$ or $o(d,1)$
 \beq
d{W_0}^{A,B}+{W_0}^{A,}_{\phantom{A,}C}
{W_0}^{C,B}=0\label{OmegaAds}. \eeq For definiteness, let us
consider the $AdS_d$ case of $o(d-1,2)$. The Lorentz connection
and frame 1-form are identified  as \beq {W_0}^{a,b} =
\omega^{a,b} \,,\qquad {W_0}^{a,\bullet} = \lambda e^a \,.\eeq
$\bullet$ denotes the extra time value of the \grAdS\ vector index
compared to the \grLorenz\ vector index. (For more detail see
Appendix A.) The dimensionful  constant $\lambda$ to be identified
with the inverse $(A)dS$ radius is introduced
 to make the vielbein dimensionless. The curvature $R=d{W_0}+{{W_0}}^2$
decomposes as $R=T^a P_a +\frac12 R^{a,b}M_{a,b}$, where the
\grAdS  generators $P_a$ and $M_{a,b}$ are  associated with \AdS\
``translations'' and Lorentz rotations, respectively. $T^a$ and
$R^{a,b}$ are the torsion tensor (\ref{ZeroTorsionA}) and the
modified Riemann tensor (\ref{ZeroCurvatureA}), respectively.
 The equation (\ref{OmegaAds})   can be rewritten as the
nilpotency condition ${D_0}^2=0$ for $D_0=d+{W_0}$.

As shown by MacDowell and Mansouri \cite{MacDowell:1977jt}, modulo
the topological Gauss-Bonnet term, the action\footnote{the
coefficient $\lambda^2$ instead of $\lambda$ is due to a somewhat
different dimension of fields describing background space in
comparison with dynamical ones. $\kappa^2$ is the gravitational
constant.} \beq S=\frac{1}{4\kappa^2\lambda^2}\int_{{\textit{M}}^d}
R^{a_1,a_2}R^{a_3,a_4}e^{a_5}\ldots e^{a_d}\epsilon_{a_1 a_2 \ldots
a_d} \eeq in $d=4$ describes  Einstein gravity with the cosmological
term. The case of arbitrary $d$ was discussed in
\cite{Vasiliev:2001wa}, where it was shown in particular that this
action still describes a massless spin two field  at the linearized
level although differs from the pure Einstein action with the
cosmological term  at the nonlinear level (for more detail see also
\cite{Bekaert:2005vh}).

A covariant generalization of the MacDowell-Mansouri approach
proposed by Stelle and West \cite{Stelle:1979aj} is achieved by
introducing a compensator field $V^A(x)$ that has  a fixed norm
\beq V^A V_A=\pm 1\label{Norm} \eeq and carries no physical
degrees of freedom. A sign ($+$)$-$  corresponds to the $(A)dS$
case. The Lorentz algebra $o(d-1,1)\subset o(d-1,2)(o(d,1))$ is
identified with the stability subalgebra of the compensator. The
vielbein and spin connection admit the covariant definition
\beq\lambda E^A=DV^A=dV^A+{W_0}^{A,}_{\phantom{A,} B} V^B
\label{EDefinition}, \eeq \beq
{{W_0}_L}^{A,B}={W_0}^{A,B}-\lambda(E^A V^B-E^B
V^A).\label{SpinConnection}\eeq As a consequence of
(\ref{EDefinition}) and (\ref{Norm}) the vielbein is orthogonal to
the compensator \beq E^A V_A=0\eeq and the Lorentz covariant
derivative of the compensator with respect to the spin connection
(\ref{SpinConnection}) is zero  \beq D_L
V^A=dV^A+{{W_0}_L}^{A,}_{\phantom{A,}B} V^B=0\,.\eeq A convenient
gauge choice, referred to as {\it standard gauge}, which breaks
local \grAdS\ $(o(d,1))$  down to the local Lorentz algebra, is
 \beq V^A=\delta^{A}_{\bullet}\eeq with the
compensator vector pointing along the extra $(d+1)^{th}$
direction. In the standard gauge, the nonzero components $E^a$ of
$E^A$  are identified with the the vielbein $e^a$.

The idea of the approach to the description of HS gauge fields
proposed in \cite{Lopatin:1987hz, Vasiliev:2001wa} was to replace
the \AdS\ isometry algebra $h=$ \grAdS\ or $o(d,1)$ by some its
extension $g$ called HS symmetry algebra in such a way that the
dynamics of massless HS fields is described in terms of the
connection 1-forms of $g$ similarly to as gravity is described by
the connections of $o(d,1)$ ($o(d-1,2)$). Note that $g$ turns out
to be infinite-dimensional.

The free field approximation results from the following
perturbative expansion. A zero-order vacuum (background) field
$\omega_0$ takes values in $h\subset g$,
 i.e., it
identifies with  the 1-form $W^{A,B}_0=-W^{B,A}_0$. It is required
to satisfy the zero-curvature equation for $h$ thus  describing
\AdS\ space-time (or Minkowski in the flat limit).

 $\omega_1$ describes  first-order fluctuations of the fields of all
spins including gravity. The HS field strength is \beq R=R_0+R_1
+o(\omega_1)\,, \eeq where \beq
R_0=d\omega_0+[\omega_0\,,\omega_0], \eeq \beq R_1= d\omega_1+
[\omega_0\,,\omega_1 ]\,. \eeq Here $[\,,]$ is the Lie product of
$g$.
 Since  $h\subset g$, the zero-order part of the
 curvature of the full HS algebra $g$ is  zero
$R_0=0$. The first-order part is  $R_1=D_0\omega_1$, where $D_0$
is the background covariant derivative built of the connection
$\omega_0$. Since ${D_0}^2=0$, $R_1$ is invariant under the
linearized gauge transformations \beq \delta \omega_1=D_0\xi\,,
\eeq where $\xi$ is an arbitrary gauge parameter that takes values
in $g$. This is the linearized HS gauge transformation. Any action
constructed in terms of $R_1$ is therefore automatically HS gauge
invariant at the linearized level.

In the known cases, $g$ decomposes into an infinite direct sum of
finite-dimensional $h$-modules under the adjoint action of
$h\subset g$. Gauge fields $\omega_1^s$, corresponding to
different irreducible submodules labeled by the integral weights
$s$ of $h$, describe fields of different ``spins'' $s$. In other
words, the connections  $\omega_1^s$ take values in irreducible
tensor $h$-modules.

As an example, a spin-$s$ symmetric massless field in  \AdS\ can
be described \cite{Vasiliev:2001wa} by the 1-form
$W^{A(s-1),B(s-1)}_{\underline{m}}$ with tangent indices that
possess the symmetry of the two-row traceless \AdS\ Young tableau
 \footnote{I.e. the field
${W_{\underline{m}}}^{A(s-1),B(s-1)}$ is symmetric in indices $A$
and $B$ separately, traceless and the symmetrization over any $s$
indices gives zero: ${W_{\underline{m}}}^{A(s-1),A B(s-2)}\equiv
0$.} \YoungTwoRowBold{5}{s-1}{5}{s-1}. (To distinguish between
Young tableaux of \grLorenz\ and \grAdS\, the latter are drown in
bold.)
 The gauge transformation law is
\beq \delta W^{A(s-1),B(s-1)}_{\{1\}}=D_0
\xi^{A(s-1),B(s-1)}_{\{0\}}.\eeq The field strength \beq
R^{A(s-1),B(s-1)}_{\{2\}}=D_0 W^{A(s-1),B(s-1)}_{\{1\}}\eeq is
gauge invariant by virtue of ${D_0}^2=0$.

To make contact with the analysis of \cite{Lopatin:1987hz}
explained in Section \ref{AdSHS} one observes that the $h$-module
$\YoungTwoRowBold{4}{s-1}{4}{s-1}$ decomposes
 (see Appendix B) into the set of $o(d-1,1)$-modules
 \beq
\begin{matrix} \YoungTwoRow{5}{s-1}{3}{k}
\end{matrix}\ ,\ k=0,1,...,(s-1),\eeq
i.e.,
$W^{A(s-1),B(s-1)}_{\{1\}}\longleftrightarrow\omega^{a(s-1),b(k)},\
k=0,1,...,(s-1)$. In particular, the vielbein-like field and the
Lorentz connection-like  auxiliary field are \beq \label{lvi}
{\omega_{\underline{m}}}^{a(s-1)}={W_{\underline{m}}}^{a(s-1),C(s-1)}V_{C_1}
\ldots V_{C_{s-1}}\,, \qquad
{\omega_{\underline{m}}}^{a(s-1),b}={W_{\underline{m}}}^{a(s-1),bC(s-1)}V_{C_1}
\ldots V_{C_{s-2}}\,. \eeq

In the \grAdS covariant approach, a general $P$-even HS gauge
invariant action is\bee \label{HigherSpinAction}\begin{split}
S^s=&\frac1{2\lambda}\sum_{q=0}^{q=s-2}b^s_q \int_{M^d}
\epsilon_{A_1 \ldots
A_{d+1}}\ V^{A_{5}} E^{A_{6}}\ldots E^{A_{d+1}}\times \\
\times &{R_{\{2\}}}^{A_1 B(s-2),A_2 C(q)D(s-q-2)} {R_{\{2\}}}^{A_3
}{}_{B(s-2)}{}^{,A_4 C(q)}{}_{D(s-q-2)}V_{C(2q)} ,\end{split} \eee
where $V_{C(p)}= \underbrace{V_C\ldots V_C}_p$. The coefficients
$b^s_q$ are fixed \cite{Vasiliev:2001wa} modulo an overall factor
\beq \label{cmas} b^s_{q}=b^{s}\frac{(d-5+2q)!!(q+1)}{q!}\eeq
 by the condition that the variation with respect to the fields
contracted with less than $s-2$ compensators is identically zero
\beq \label{HS_EFDC_ADS} \Pi_L \left [V^{C(k)}\frac{\delta
S}{\delta W^{A(s-1),B(s-k-1)C(k)}}\right ]\equiv0,\qquad k<s-2\,,
\eeq where $\Pi_L$ is the projector to the $V^A$- transversal part
of a tensor. By virtue of (\ref{lvi}), this condition is
equivalent to the extra field decoupling condition
(\ref{HS_EFDC_Lorentz}). As a result, modulo total derivatives,
the action depends only on the vielbein-like and Lorentz
connection-like fields (\ref{lvi}), being free of higher
derivatives.

Now we are in a position to apply  this approach to the
description of partially massless fields. For more detail on the
HS gauge theory we refer the reader to \cite{Bekaert:2005vh,rev}.

\section{Partially Massless Higher-Spin Fields in the Frame-Like
Formalism}\setcounter{equation}{0} \label{ResultsReview}
\subsection{Summary of results}
Our main result is that a free partially massless field of
spin-$s$ and depth-$t$ can be described by a gauge field
$W^{A(s-1),B(t)}_{\{1\}}$, which has the symmetry of two-row
traceless $o(d,1)$ or \grAdS\ Young tableau \footnote{I.e.
$W^{A(s-1),B(t)}$ is a traceless tensor  symmetric  in indices $A$
and $B$ separately, that satisfies  the antisymmetry property
$W^{A(s-1),AB(t-1)}\equiv0$. } \YoungTwoRowBold{5}{s-1}{3}{t} with
a gauge transformation law \bee \delta W_{\{1\}}^{A(s-1),B(t)}=D_0
\xi_{\{0\}}^{A(s-1),B(t)}
 \eee
and a field strength
 \bee \label{pstrength} R_{\{2\}}^{A(s-1),B(t)}=D_0
W^{A(s-1),B(t)}_{\{1\}}\,. \eee Since ${D_0}^2=0$, the field
strength  is gauge invariant  and satisfies  Bianchi identities
\beq\label{GeneralBianchiIden} \delta
R_{\{2\}}^{A(s-1),B(t)}=0,\qquad D_0 R_{\{2\}}^{A(s-1),B(t)}=0.
\eeq

In the rest of this section we sketch the main results of our
analysis leaving some details of their derivation to a later
publication
 \cite{Skvortsov:2006}, where it will be shown in particular,
 that all gauge symmetry parameters $\xi_{\{0\}}^{A(s-1),B(t)}$ are
Stueckelberg (i.e., analogous to the Lorentz symmetry parameters
in the case of gravity) except for the parameter \beq
\xi^{b(t)}=\xi_{\{0\}}^{A(s-1),b(t)}V_{A_1}\ldots V_{A_{s-1}} \,,
\eeq the most $V$-tangential part of $\xi_{\{0\}}^{A(s-1),B(t)}$,
which is traceless as a consequence of the tracelessness and the
symmetry properties of $\xi_{\{0\}}^{A(s-1),B(t)}$ analogous to
those of $W_{\{1\}}^{A(s-1),B(t)}$. Indeed, in the standard gauge
$V^A=\delta^A_{\bullet}$ \beq \label{autt}
\xi^{b(t)}\eta_{bb}=-\eta_{\bullet\bullet}\xi_{\{0\}}^{A(s-1),\bullet\bullet
b(t-2)}V_{A_1}\ldots
V_{A_{s-1}}=\xi_{\{0\}}^{{\bullet\ldots\bullet},\bullet\bullet
b(t-2)}\equiv0,\eeq since the symmetrization over any $s$ indices
of $\xi_{\{0\}}^{A(s-1),B(t)}$ gives zero.

By dynamical fields we mean those which are neither pure gauge
(i.e., Stueckelberg) nor auxiliary (i.e., to be expressed in terms
of derivatives of the other fields by virtue of constraints). The
dynamical fields
 in the system with $t<s-1$ are \cite{Skvortsov:2006}
 traceless symmetric tensors $\phi^{a(s)}$,
$\phi^{a(t)}$ and $\phi^{a(t-1)}$ (the latter field is absent for
$t=0$) defined as follows

\beq \phi^{a(s)}=\traceless{W^{a(s-1),C(t)|a}V_{C_1}\ldots
V_{C_t}}\,, \eeq where $\traceless{}$ is a projector to the
traceless part, \beq \phi^{a(t-1)}= {W_m}^{ma(t-1) C(s-t-1),C(t)}
V_{C_1}\ldots V_{C_{s-1}},\qquad t>0\eeq and $\phi^{a(t)}$ is a
linear combination of the form \beq \phi^{a(t)}=\alpha {W_m}^{m
a(t) C(s-t-2),C(t)} V_{C_1}\ldots V_{C_{s-2}}\ + \beta
{W_m}^{a(t)C(s-t-1),C(t-1)m} V_{C_1}\ldots V_{C_{s-2}}. \eeq Note
that the fields $\phi^{a(t)}$ and $\phi^{a(t-1)}$ are
automatically  traceless  (cf. (\ref{autt})).

The gauge transformation law for the field $\phi^{a(s)}$ is \beq
\delta \phi^{a(s)}=\traceless{\overbrace{D^a\ldots D^a}^{s-t}
\xi^{a(t)}}\,. \eeq It coincides with the transformation law
(\ref{gtr}) for partially massless fields found  by Deser and
Waldron \cite{DeserWaldron} and Zinoviev \cite{Zinoviev}.

Gauge invariant combinations of derivatives of the dynamical
fields that are non-zero on-shell we call generalized Weyl
tensors. In practice, to find generalized Weyl tensors, one should
set to zero as many components of the gauge invariant field
strengths as possible by imposing constraints that express
auxiliary fields in terms of derivatives of the dynamical fields
(these generalize the zero-torsion condition in  gravity). Other
components of the gauge invariant field strengths are zero by
virtue of the field equations (these generalize the Einstein
equations in gravity). Some more vanish by virtue of Bianchi
identities applied to the former two (e.g., the cyclic part of the
Riemann tensor in gravity is zero by virtue of  the Bianchi
identity applied to the zero-torsion condition). What is left
nonzero are the generalized Weyl tensors.

In the massless case with $t=s-1$, the generalized Weyl tensor is
known \cite{Vasiliev:2001wa} to be described by the length $s$
two-row rectangular $o(d-1,1)$ Young tableau. In the partially
massless case, a generalized Weyl tensor
 identifies \cite{Skvortsov:2006}
with the traceless component of  $R^{a(s-1),b(t)}_{\{2\}}$ (which
is the $V^A$ orthogonal part of  $R_{\{2\}}^{A(s-1),B(t)}$), which
has the symmetry property of \YoungTwoRow{5}{s}{4}{t+1}.

The case of $t=0$, which corresponds to the conformal
generalization of the Maxwell theory discussed in
\cite{DeserWaldron}, is special.  The $t=0$ generalized Weyl
tensors are the components of ${R_{\n{m}\n{n}}}^{a(s-1)}$ with the
symmetry properties of the Young tableaux \YoungTwoRow{5}{s}{1}{}
and $\YoungOneRow{4}{s-1}\,$ for $s>1$.

The geometric formulation of the partially massless fields will be
illustrated in Section \ref{Samples} by the  examples of $s=2$,
$t=0$ with the gauge field ${W_{\n{m}}}^{A}$ and $s=3$, $t=1$ with
the gauge field ${W_{\n{m}}}^{AA,B}$. In the spin-2 example, the
dynamical fields $\phi^{a(s)}$, $\phi^{a(t)}$ correspond to
$\traceless{W^{(a|a)}}$ and $W^{m|}_{\phantom{m|}m}$ while
$\phi^{a(t-1)}$ is absent as $t=0$. In the spin-3 example
$\phi^{a(s=3)}=\traceless{W^{(aa|a)}}$, $\phi^{a(t=1)}=\alpha
W^{am,\bullet|}_{\phantom{am,\bullet|}m}+\beta
W^{[a\bullet,m]|}_{\phantom{[a\bullet,m]|}m}$ and
$\phi^{a(t-1=0)}=W^{m\bullet,\bullet|}_{\phantom{m\bullet,\bullet|}m}$.

\subsection{\AdS covariant action}

Free gauge invariant action functionals are easily constructed in
terms of the gauge invariant linearized curvatures in the form
 \bee
\begin{split}& S^{s,t}=\frac1{2\lambda} \sum_{k,m} \tilde{b}^{s,t}_{k,m}
\int_{M^d}   \epsilon_{A_1 A_2 \ldots A_{d+1}} V^{A_5}
E^{A_6}\ldots E^{A_{d+1}} \\ & {R_{\{2\}}}^{A_1 B(s-k-2)C(k),A_2
D(t-m-1)C(m)}  {R_{\{2\}}}^{A_3 \phantom{B(s-k-2)}C(k),A_4
\phantom{D(t-m-1)}C(m)}_{\phantom{A_3} B(s-k-2)\phantom{C(k),A_4}
D(t-m-1)} V_{C(2k+2m)}\label{PartAction}\,,\end{split} \eee where
\beq \tilde{b}^{s,t}_{k,m}=b^{s,t}_{k,m}
\theta(m)\theta(k)\theta(t-m-1)\theta(s-m-k-2)\eeq with
\beq\begin{cases}
\theta(n) =1\qquad n\geq 0,\\
\theta(n) =0 \qquad n<0\,.\end{cases} \eeq implies that the
summation range is finite for a fixed spin.

The parameters $s$ and $t$, which determine a field type, are
fixed. Different choices of the coefficients $b^{s,t}_{k,m}$ lead
to different gauge invariant dynamical systems. To simplify
formulae, the range of summation over the indices $k$ and $m$ in
(\ref{PartAction}) is chosen so that the terms with $k>s-t-1$ are
not independent, being expressible via linear combinations of
terms with smaller $k$ because the symmetrization over any $s$
tangent indices of the HS curvatures gives zero.

Generally, different gauge connections that enter the curvatures
in (\ref{PartAction}) are independent fields, which may carry
their own degrees of freedom. Alternatively, all connections can
be expressed in terms of (derivatives of) certain  dynamical
fields by virtue of imposing constraints that generalize the zero
torsion condition in gravity and require some components of the
field strengths to vanish. Plugging the resulting expressions back
into the action gives rise to a gauge invariant higher derivative
action, which  may again describe a system with extra degrees of
freedom. To avoid these unwanted degrees of freedom, the
coefficients in the action (\ref{PartAction}) should be adjusted
so that, in the flat limit, the action should amount to the sum of
free massless actions for spins from $t+1$ to $s$ in agreement
with the construction of Zinoviev discussed in Section
\ref{ZinovievApproach}. It turns out that for the models with
$t>1$ this is achieved by imposing the extra field decoupling
condition
 \bee
 \label{PM_EFDC_ADS} \Pi_L \left [V^{C(m+k)}\frac{\delta
S^{s,t}}{\delta W^{A(s-k-1)C(k),B(t-m)C(m)}}\right ]\equiv0,\qquad
\begin{tabular}{l}
 $m=0,1,...,t-2,$ \\
 $k=0,1,...,s-t-1$ \\
\end{tabular},
\eee which means that the variation is  nonzero only for the
components $W^{A(s-1),B C(t-1)}V_{C(t-1)}$ and generalizes
(\ref{HS_EFDC_ADS}) to partially massless fields.

 Using
\bee \delta {R_{\{2\}}}^{A(s-1),B(t)}=D_0 \delta
{W_{\{1\}}}^{A(s-1),B(t)}\,, \eee the symmetry properties of the
field strength, the Bianchi identities (\ref{GeneralBianchiIden}),
the definition of the frame (\ref{EDefinition}) along with its
consequence $D_0 E^A=0 $ and the identities \bee \epsilon_{A_1
\ldots A_{d+1}}=V^F(V_{A_1}\epsilon_{F A_2 \ldots
A_{d+1}}&+V_{A_2}\epsilon_{A_1 F A_3 \ldots A_{d+1}}+\ldots+
V_{A_{d+1}}\epsilon_{A_1 \ldots A_{d} F}),\eee

\bee \begin{split} \epsilon_{ A_1 \ldots A_5 H_6 \ldots
H_{d+1}}& E^F E^{H_6} \ldots E^{H_{d+1}}= \\
&\frac1{d-3}\left(\sum^{k=5}_{k=1} \epsilon_{A_1 \ldots
\widehat{\textbf{A}}_k \ldots A_5 H_5 \ldots
H_{d+1}}\delta^F_{A_k}(-)^{k+1}\right)E^{H_5}\ldots
E^{H_{d+1}}\,,\mathrm{}
\end{split}\eee
which manifest the fact that the antisymmetrization of $d+2$
indices, which take $d+1$ values, is zero, we obtain \bee
\label{PMActionVariationReduced}
\begin{split}  &\delta S^{s,t}=\frac1{d-3}\sum_{k,m}\int_{M^d}
\epsilon_{A_1 A_2 \ldots
A_{d+1}} V^{A_4} E^{A_5}\ldots E^{A_{d+1}}V_{C(2k+2m+1)}\times   \\
\times & \{\delta W^{A_1 B(s-k-2) C(k),A_2 D(t-m-1) C(m)}  R^{A_3
\phantom{B(s-k-2)}C(k), \phantom{D(t-m-1)}C(m+1)}_{\phantom{A_3}
B(s-k-2)\phantom{C(k),} D(t-m-1)}\times \\
\times & \left(\left[d-3+2(k+m)\right]\tilde{b}^{s,t}_{k,m}-(m+1)
\tilde{b}^{s,t}_{k,m+1}+\frac{(k+1)^2}{(s-k-2)}\tilde{b}^{s,t}_{k+1,m}\right)+
\\+& \delta W^{A_1 B(s-k-2) C(k),A_2 D(t-m-1) C(m)}
R^{\phantom{B(s-k-2)}C(k+1),A_3
\phantom{D(t-m-1)}C(m)}_{B(s-k-2)\phantom{C(k+1),A_3} D(t-m-1)}
\times \\ \times&
\left(-\left[d-3+2(k+m)\right]\tilde{b}^{s,t}_{k,m}+
\frac{(k+1)(s-k-m-1)}{(s-k-2)}\tilde{b}^{s,t}_{k+1,m}\right)+
\\
&+ (\delta W \longleftrightarrow R) \}\,.
\end{split} \eee

 The extra field decoupling condition (\ref{PM_EFDC_ADS}) demands
 most of the terms in the variation (\ref{PMActionVariationReduced})
 to vanish. This is achieved by virtue of imposing the following conditions
 on the coefficients
\beq \label{recurrsysa} \begin{cases} \displaystyle
-\left[d-3+2(k+m)\right]
b^{s,t}_{k,m}+\frac{(k+1)(s-k-m-1)}{(s-k-2)}b^{s,t}_{k+1,m}=0, \\
\displaystyle
\left[d-3+2(k+m)\right]b^{s,t}_{k,m}-(m+1)b^{s,t}_{k,m+1}+
\frac{(k+1)^2}{(s-k-2)}b^{s,t}_{k+1,m}=0,
\\
t>1,\ k=0,...,s-t-1,\ m=0,...,t-2\,,
\end{cases} \eeq
which fix $b^{s,t}_{k,m}$ up to an overall factor $b^{s,t}$ \beq
\label{Coefficientsb} b^{s,t}_{k,m}=b^{s,t}
\frac{(s-k-m-1)!(d-5+2(k+m))!!}{k!m!(s-k-2)!(s-m)!}. \eeq

The remaining  nonzero part of the variation of the action is \bee
\begin{split}  &\delta S^{s,t}=\frac1{d-3}\sum_k\int_{M^d}
\epsilon_{A_1 A_2 \ldots A_{d+1}} V^{A_4} E^{A_5}\ldots
E^{A_{d+1}}V_{C(2k+2t-1)}\times \\ \times & \{\delta W^{A_1
B(s-k-2) C(k),A_2 C(t-1)}  R^{A_3 \phantom{B(s-k-2)}C(k),
C(t)}_{\phantom{A_3} B(s-k-2)\phantom{C(k),}} + (\delta W
\longleftrightarrow R) \}\times\\ \times &
\left(\left[d-5+2(k+t)\right]\tilde{b}^{s,t}_{k,t-1}+
\frac{(k+1)^2}{(s-k-2)}\tilde{b}^{s,t}_{k+1,t-1}\right)\,.
\end{split} \eee

 As explained in Section
\ref{LorentzCovariantAction}, the coefficients
(\ref{Coefficientsb}) guarantee the consistency of the flat limit,
which is in agreement with the construction of Zinoviev. For the
massless case of $t=s-1$, the variation reduces to that of the
action (\ref{HigherSpinAction}) \cite{Vasiliev:2001wa}.

The partially massless cases  with $t=0$ and $t=1$ are special. In
the case of $t=1$, the variation of the action is
 \bee \label{PM_Variation_t_one}
\begin{split} &\delta S^{s,1} =\frac1{d-3}
\sum_{k} \int_{M^d}  \epsilon_{A_1 A_2 \ldots A_{d+1}} V^{A_4}
E^{A_5}\ldots E^{A_{d+1}}V_{C(2k+1)}\{ \\ & \delta W^{A_1 B(s-k-2)
C(k),A_2} R^{A_3 \phantom{B(s-2-k)}C(k),C}_{\phantom{A_3}
B(s-k-2)}
[(d-3+2k)\tilde{b}^{s,1}_{k,0}+\frac{(k+1)^2}{(s-k-2)}\tilde{b}^{s,1}_{k+1,0}]
+
\\+&\delta W^{A_1 B(s-k-2) C(k),A_2}
R^{\phantom{B(s-k-2)}C(k+1),A_3}_{B(s-k-2)}
[-(d-3+2k)\tilde{b}^{s,1}_{k,0}+\frac{(k+1)(s-k-1)}{(s-k-2)}
\tilde{b}^{s,1}_{k+1,0}] +\\+&(\delta W \longleftrightarrow R )\}.
\end{split}.\eee
The extra fields decoupling condition (\ref{PM_EFDC_ADS}) is not
applicable for this case and the coefficients are fixed directly
from the requirement of consistency of the flat limit. As
explained in Subsection \ref{LorentzCovariantAction},
  upon an appropriate rescaling, the first term in
(\ref{PM_Variation_t_one}) tends in the flat limit to the sum of
variations of massless actions for spins from  $(t+1)$ to $s$,
while the second term diverges. The correct flat limit therefore
requires the second term  to be zero that fixes $b^{s,t}_{k,0}$ in
the form \beq b^{s,1}_{k,0}=b^{s,1}\frac{(s-k-1)(d-5+2k)!!}{k!s!},
\eeq i.e., (\ref{Coefficientsb}) is still true.

 In the case of $t=0$, the action cannot be  written at
all  in the form (\ref{PartAction}) because there is no room for
the indices $A_2$ and $A_4$. For this case the action is
constructed in terms of the generalized Weyl tensor $C^{a(s),b}$
\beq \label{ActionWeyl} S^{s,0}=\frac12\int_{M^d}
\epsilon_{c_1\ldots c_d}e^{c_1}\ldots e^{c_d}
\left(C^{a(s),b}C_{a(s),b}-\frac12
C^{a(s-1)m,}_{\phantom{a(s-1)m,}m}C_{a(s-1)n,}^{\phantom{a(s-1)n,}n}\right)
.\eeq Note that, in $d=4$, this case corresponds to the conformal
partially massless fields of Deser and Waldron \cite{DeserWaldron}
with $4d$ Maxwell electrodynamics as a particular case of $s=1$,
where $C^{a,b}$ identifies with the Maxwell field strength. The
form of the action (\ref{ActionWeyl}) is in agreement with the
well-known fact that the  Maxwell lagrangian 4-form cannot be
written as the exterior product of the Maxwell 2-form field
strengths.

\subsection{Lorentz covariant formulation} \label{LorentzCovariantAction}
The practical analysis of the field content of the model under
consideration is more illuminating in terms of the Lorentz
irreducible components of the fields. An irrep of \grAdS(o(d,1))
with the Young symmetry type \YoungTwoRowBold{6}{s-1}{4}{t}
decomposes into the following set of irreps of \grLorenz \beq
\YoungTwoRowBold{6}{s-1}{4}{t} = \bigoplus_{k=t,\  l=0}^{k=s-1,\
l=t}\quad \YoungTwoRow{5}{k}{3}{l},\qquad \eeq i.e., in terms of
Lorentz tensors, the partially massless theory of spin-$s$ and
depth-$t$ is described by 1-forms
 $\omega_{\{1\}}^{a(k),b(l)}\equiv \omega_{\{1\}}^{\{k,\ l\}}$\ with $k=t,...,s-1$,
$l=0,1,...,t$.

The field strengths have the following structure \bee
\label{LorentzCovariantFieldStrengths}
\begin{split} R_{\{2\}}^{\{k,\ l\}}=D_0 (\omega_{\{1\}}^{\{k,\ l\}})=&D
(\omega_{\{1\}}^{\{k,\ l\}}) + \lambda\sigAm
(\omega_{\{1\}}^{\{k+1,\ l\}})+ \lambda\sigBm
(\omega_{\{1\}}^{\{k,\ l+1\}})
+\\+&\lambda\sigAp(\omega_{\{1\}}^{\{k-1,\
l\}})+\lambda\sigBp(\omega_{\{1\}}^{\{k,\ l-1\}}).\end{split}\eee
The operators $\sigma^{1,2}_{\pm}$ have the form \beq
\label{sigAp}
\begin{split}
\sigAp(C^{a(k),b(l)})&=g(k,l)(k+1)(e^{a}_{}C^{a(k),b(l)}-
\frac{k}{2(d+2k-2)}e_{m}C^{a(k-1)m,b(l)}\eta^{aa}-\\
&-\frac{l}{d+k+l-3}e_{m}C^{a(k),mb(l-1)}\eta^{ab}+\\&+
\frac{k}{l(d+2k-2)(d+k+l-3)}e_{m}C^{a(k-1)b,b(l-1)m}\eta^{aa}),
\end{split} \eeq

\beq \label{sigBp}\begin{split}
\sigBp(C^{a(k),b(l)})&=G(k,l)(l+1)(e^{b}_{}C^{a(k),b(l)}-
\frac{k}{k-l}e^{a}_{}C^{a(k-1)b,b(l)}+\\&-
\frac{l}{d+2l-4}e_{m}C^{a(k),b(l-1)m}\eta^{bb}+\\&-
\frac{k(k-l-1)}{(k-l)(d+k+l-3)}e_{m}C^{a(k-1)m,b(l)}\eta^{ab}+\\&+\frac{l
k(d+2k-4)}{(k-l)(d+2l-4)(d+k+l-3)}e_{m}C^{a(k-1)b,b(l-1)m}\eta^{ab}+\\&+
\frac{k(k-1)}{(k-l)(d+k+l-3)}e_{m}C^{a(k-2)mb,b(l)}\eta^{aa}+\\&-\frac{l
k(k-1)}{(k-l)(d+k+l-3)(d+2l-4)}e_{m}C^{a(k-2)bb,b(l-1)m}\eta^{aa})
,\end{split}\eeq \beq
\label{sigAm}\sigAm(C^{a(k),b(l)})=f(k,l)(e_{m}C^{a(k-1)m,b(l)}
+\frac{l}{k-l+1}e_{m}C^{a(k-1)b,b(l-1)m}), \eeq \beq \label{sigBm}
\sigBm(C^{a(k),b(l)})=F(k,l)(e_{m}C^{a(k),b(l-1)m})\,,\eeq where
$g(k,l)$, $G(k,l)$, $f(k,l)$ and $F(k,l)$ are some coefficients,
which can be either derived from the covariant field strengths
(\ref{pstrength}) or obtained directly from the condition of the
gauge invariance of $R_{\{2\}}^{a(k),b(l)}$. Their explicit form
is given in Appendix C. (For more detail see
\cite{Skvortsov:2006}).

A general $P$-even $o(d-1,1)$-covariant gauge invariant action has
the following form in terms of Lorentz covariant field strengths
  \bee \label{Lorenzcov} S^{s,t}= \frac{1}{2\lambda}
\sum_{k,m} \tilde{a}^{s,t}_{k,m} \int_{M^d} \epsilon_{c_1 \ldots
c_d} e^{c_5} \ldots e^{c_d}
 {R_{\{2\}}}^{c_1 a(k-1),c_2 b(m-1)}
{R_{\{2\}}}^{c_3\phantom{a(k-1)},c_4
}_{\phantom{c_3}a(k-1)\phantom{,c_4}b(m-1)}, \eee where
\beq\tilde{a}^{s,t}_{k,m}=a^{s,t}_{k,m}\theta(k-t)\theta(m)\theta(t-m)\theta(s-k-1)
\eeq and the coefficients $a^{s,t}_{k,m}$ remain to be determined.
This lagrangian generalizes to partially massless fields that
introduced in \cite{Lopatin:1987hz} for  massless fields
($t=s-1$).

Let us introduce the inner product \beq
\label{MyScalarProduct}\langle\phi^{\{k,l\}}_{\{p\}} |
\psi^{\{k,l\}}_{\{q\}} \rangle = \int_{M^d} \epsilon_{c_1 c_2
\ldots c_d}e^{c_5}\wedge \ldots \wedge e^{c_d} \wedge \phi^{c_1
a(k-1),c_2 b(l-1)}_{\{p\}} \wedge \psi^{c_3\phantom{a(k-1)}\
,c_4}_{ {\{q\}} a(k-1)\phantom{\{q\},}b(l-1)}\,, \eeq which is
nonzero if the  $p$-form $\phi^{a(k), b(l)}_{\{p\}}$ and $q$-form
$\psi^{a(k), b(l)}_{\{q\}}$
 carry equivalent representations of the Lorentz group
 (in particular, have the same symmetry type) and are such that $p+q=4$.
Then the action (\ref{Lorenzcov}) reads
  \beq S^{s,t}=
\frac{1}{2\lambda} \sum_{k,m} \tilde{a}^{s,t}_{k,m} \langle
R_{\{2\}}^{\{k,m\}}|R_{\{2\}}^{\{k,m\}}\rangle.\eeq

The normalization coefficients $g_{k,l}$, $G_{k,l}$, $f_{k,l}$ and
$F_{k,l}$  are chosen so (see Appendix C) that the operators
$\sigma^{1,2}_{+}$ are conjugated to $\sigma^{1,2}_{-}$ with
respect to this product \beq \langle\phi^{\{k,l\}}_{\{p\}}
|\sigma^{1,2}_{+}| \psi^{\{k,l\}}_{\{q\}} \rangle = (-)^{pq+p+q}
\langle\psi^{\{k,l\}}_{\{q\}} |\sigma^{1,2}_{-}|
\phi^{\{k,l\}}_{\{p\}} \rangle \,.\eeq Making use of the Bianchi
identities along with \beq
\begin{split} \delta R_{\{2\}}^{\{k,m\}}=D_0 (\delta
\omega_{\{1\}}^{\{k,m\}})=&D(\delta\omega_{\{1\}}^{\{k,m\}})+\lambda\sigAm
(\delta\omega_{\{1\}}^{\{k+1,m\}})+\lambda\sigBm(\delta\omega_{\{1\}}^{\{k,m+1\}})+
\\+&\lambda\sigAp(\delta\omega_{\{1\}}^{\{k-1,m\}})+\lambda\sigBp(\delta
\omega_{\{1\}}^{\{k,m-1\}})\end{split}\eeq
 the variation of the action amounts to
\beq
\begin{split}\delta S^{s,t} =
\sum_{k,m}&\langle \delta
\omega^{\{k,m\}}|\sigAm|R^{\{k+1,m\}}\rangle
(\tilde{a}^{s,t}_{k+1,m}-\tilde{a}^{s,t}_{k,m})+\\+&\langle \delta
\omega^{\{k,m\}}|\sigBm|R^{\{k,m+1\}}\rangle
(\tilde{a}^{s,t}_{k,m+1}-\tilde{a}^{s,t}_{k,m})+\\+&\langle \delta
\omega^{\{k,m\}}|\sigAp|R^{\{k-1,m\}}\rangle
(\tilde{a}^{s,t}_{k-1,m}-\tilde{a}^{s,t}_{k,m})+\\+&\langle \delta
\omega^{\{k,m\}}|\sigBp|R^{\{k,m-1\}}\rangle
(\tilde{a}^{s,t}_{k,m-1}-\tilde{a}^{s,t}_{k,m}).
\end{split} \eeq

The 1-form $W^{A(s-1),B(t)}$  is equivalent to the set of Lorentz
1-forms $\omega^{a(l),b(u)}$ with $l=t,\ldots s-1$ and
$u=0,1\ldots t$. It is easy to see that
 $W^{A(s-1),BC(t-1)}V_{C(t-1)}$ is
 equivalent to $\omega^{a(l)}$
and $\omega^{a(l),b}$ with $l=t,\ldots s-1$. As a result, in terms
of Lorentz tensors, the extra field decoupling condition
(\ref{PM_EFDC_ADS}) takes the following simple form \beq
\label{PM_EFDC_Lorentz} \frac{\delta S^{s,t}}{\delta
\omega^{a(k),b(m)}}\equiv0,\qquad k=t,...,s-1,\quad m=2,3,...,t.
\eeq It is satisfied with \beq \label{LorentzCoefficients}
a^{s,t}_{k,m}= a^{s,t}, \eeq For the case of $t>1$ we have \beq
\begin{split}\delta S^{s,t} = a^{s,t}\sum_{k=t}^{k=s-1}
& \left(\langle \delta
\omega^{\{k,0\}}|\sigBm|R^{\{k,1\}}\rangle-\langle \delta
\omega^{\{k,1\}}|\sigBp|R^{\{k,0\}}\rangle\right)
\end{split}.
\eeq Upon the rescaling \beq \begin{split} \label{Rescaling}
&\omega^{a(k),b}\rightarrow\frac1{{\lambda}}\omega^{a(k),b}\,,\quad
\quad R^{a(k),b}\rightarrow \frac1{{\lambda}}R^{a(k),b}\\
& S\rightarrow\lambda S
\end{split} \eeq the field strengths
(\ref{LorentzCovariantFieldStrengths}) modify to
 \beq R_{\{2\}}^{\{k,0\}}= D(\omega_{\{1\}}^{\{k,0\}}) +
 \sigAm(\omega_{\{1\}}^{\{k,1\}})+{\it O}(\lambda),\eeq \beq R_{\{2\}}^{\{k,1\}}=D (\omega_{\{1\}}^{\{k,1\}}) + {\it
O}(\lambda)\,. \eeq In the flat limit $\lambda\rightarrow0$, terms
that mix different $k$ tend to zero and the field strengths take
the form of the flat space field strengths for usual massless
fields \cite{Lopatin:1987hz} \beq R_{\{2\}}^{\{k,0\}}=
D(\omega_{\{1\}}^{\{k,0\}})+\sigAm (\omega_{\{1\}}^{\{k,1\}}),\eeq
\beq R_{\{2\}}^{\{k,1\}}=D (\omega_{\{1\}}^{\{k,1\}}). \eeq

As a result, for $t>1$, the extra field decoupling condition
(\ref{PM_EFDC_Lorentz}) guarantees that the flat limit of the
action (\ref{Lorenzcov}) is a sum of free actions for massless
fields of spins from $t+1$ to $s$ which is in agreement with the
construction of Zinoviev \cite{Zinoviev}. The totally symmetric
parts $\omega^{a(k)|a}$ of ${\omega_{\n{m}}}^{a(k)}$,
$k=t,...,s-1$ identify with the double traceless Fronsdal fields
used by Zinoviev.

As it will be explained in more detail in \cite{Skvortsov:2006},
for $\lambda \neq 0$, all fields, that are not pure gauge,
 can be expressed in terms
of derivatives of the dynamical fields $\phi^{a(s)}$,
$\phi^{a(t)}$ and $\phi^{a(t-1)}$ by virtue of constraints
${R_{\underline{mn}}}^{a(k)}=0$,
$R^{a(k)m|a}_{\phantom{a(k)m|a}m}=0$, which are particular field
equations.
 As discussed in Subsection \ref{The approach of Deser and Waldron},
within the approach of Deser and Waldron \cite{DeserWaldron}, the
constraints result from the divergences of field equations.
 Note that the reduction of a number of fields compared to those of
 the approach of Zinoviev does not imply the
 reduction of a number of degrees of freedom because, as a result of
 resolution of constrains, the action expressed in terms of the dynamical
 fields contains higher derivatives for a general partially massless system.

In the special case of $t=1$, where the  extra field decoupling
condition (\ref{PM_EFDC_Lorentz}) does not apply, the variation of
the action (\ref{Lorenzcov}) in terms of  the rescaled variables
(\ref{Rescaling}) is \beq
\begin{split}\delta S^{s,1} = \sum_{k=1}^{k=s-1} &
\left(\langle \delta
\omega^{\{k,0\}}|\sigBm|R^{\{k,1\}}\rangle-\langle \delta
\omega^{\{k,1\}}|\sigBp|R^{\{k,0\}}\rangle\right) a^{s,1}_{k,0}+\\
\ls\ls+& \frac{1}{\lambda}\left(\langle \delta
\omega^{\{k,1\}}|\sigAp|R^{\{k-1,1\}}\rangle-\langle \delta
\omega^{\{k-1,1\}}|\sigAm|R^{\{k,1\}}\rangle\right)
(\tilde{a}^{s,1}_{k-1,0}-\tilde{a}^{s,1}_{k,0})\,.
\end{split}\eeq
To get rid of the term singular in the flat limit $\lambda \to 0$,
we  set $a^{s,1}_{k,0}=a^{s,1}$ so that
(\ref{LorentzCoefficients}) is still true. The remaining variation
reduces in the flat limit to that of the sum of actions for
massless fields of spins from 2 to $ s$.

\section{Examples} \setcounter{equation}{0} \label{Samples}
In this section the proposed approach is illustrated by the
simplest examples of spin-2 and spin-3 partially massless fields.

\subsection{Spin-2}
\label{SpinTwoSample} Let us consider  the
  model with $s=2$, $t=0$. The corresponding
 gauge field is a 1-form $W^A_{\{1\}}$ with
the gauge transformation law \beq\label{SpinTwoSGaugeLaw} \delta
W_{\{1\}}^A=D_{0}\xi_{\{0\}}^A\,. \eeq The gauge invariant field
strength \beq\label{SpinTwoSStrength} R_{\{2\}}^A=D_{0}W_{\{1\}}^A
\eeq
 satisfies the
Bianchi identity \beq \label{SpinTwoSBianchi}D_0
R_{\{2\}}^A\equiv0\, . \eeq

The Lorentz reduction of the \AdS\ vector \YoungA{} gives two
Lorentz irreducible representations $
\begin{matrix} \bullet \oplus\, \YoungA{}\,
\end{matrix}$ , i.e.,
\beq\begin{split}\xi_{\{0\}}^A&
\longleftrightarrow\xi=\lambda^{-1}\xi_{\{0\}}^{\bullet},\qquad\qquad
\xi^a=\xi_{\{0\}}^a\,,\\
W^A_{\{1\}}&\longleftrightarrow
{\omega_{\underline{m}}}=\lambda^{-1} W^{\bullet}_{\{1\}},\qquad
{\omega_{\underline{m}}}^a=W^a_{\{1\}},\\
R_{\{2\}}^A&\longleftrightarrow
{R_{\underline{m}\underline{n}}}=\lambda^{-1}
R^{\bullet}_{\{2\}},\qquad
{R_{\underline{m}\underline{n}}}^a=R^a_{\{2\}}\,,
\end{split}\eeq
where the rescaling of the first component by $\lambda$ is
introduced for the future convenience. The gauge transformation
law (\ref{SpinTwoSGaugeLaw}) gives
 \beq
 \label{stu}
\delta \omega_{\underline{m}}=D_{\underline{m}}\xi-
e_{\underline{m} b}\xi^b, \eeq \beq \label{gaugeSpinTwo}
 \delta
{\omega_{\underline{m}}}^{a}=D_{\underline{m}}\xi^a+\lambda^2
{e_{\underline{m}}}^a\xi\,. \eeq {}From the form of the variation
(\ref{stu}) it follows that $\omega_{\underline{m}}$ can be gauge
fixed to zero \beq \label{gauge} \omega_{\underline{m}} = 0 \eeq
by adjusting the Stueckelberg parameter $\xi^a$. Then, from the
condition $\delta\omega_{\underline{m}}\equiv0$ it follows that
the leftover gauge symmetry with the parameter $\xi$ requires
$\xi^a= D^a\xi$. Substituting this into the gauge transformation
law (\ref{gaugeSpinTwo}) we obtain
  \beq \delta
{\omega_{\underline{m}}}^{a}=D_{\underline{m}}D^a\xi+\lambda^2
{e_{\underline{m}}}^a\xi.\eeq The field
$\omega^{a|b}={\omega_{\underline{m}}}^{a} e^{b\underline{m}}$
decomposes into $\YoungAA{15}\ \oplus\ \YoungB{}\oplus{\bullet}$,
i.e., into antisymmetric part $\YoungAA{15}=\omega^{[a|b]}$ and
traceful symmetric part $\YoungB{}\oplus{\bullet}=\omega^{(a|b)}$
with the scalar trace field $\bullet=\omega^{c|}_{\phantom{c|}c}$.
In these terms the gauge transformation law (\ref{gaugeSpinTwo})
reads \beq \delta \omega^{[a|b]}=  0,\eeq \beq
\label{SpinTwoGaugeLaw}\delta\omega^{(a|b)}=
D^{(a}D^{b)}\xi+\lambda^2 \eta^{ab}\xi \,. \eeq We see that
$\omega^{[a|b]}$ is gauge invariant while
$h^{ab}\equiv\omega^{(a|b)}$ transforms as a partially massless
spin-2 field (\ref{PMSpinTwoGaugeLaw}). The field strength
(\ref{SpinTwoSStrength}) reduces to \beq\begin{split}
R_{\underline{m}\underline{n}}&=D_{\underline{m}}\omega_{\underline{n}}-
e_{\underline{m} b}{\omega_{\underline{n}}}^{b},\\
{R_{\underline{m}\underline{n}}}^a&=D_{\underline{m}}
{\omega_{\underline{n}}}^{a}+\lambda^2
{e_{\underline{m}}}^a\omega_{\underline{n}}.\end{split}\eeq In the
gauge (\ref{gauge}) we have
 \beq\label{SpinTwoSGaugedStrengths}\begin{split}
R_{\underline{m}\underline{n}}&=-
e_{\underline{m} b}{\omega_{\underline{n}}}^{b},\\
{R_{\underline{m}\underline{n}}}^a&=D_{\underline{m}}
{\omega_{\underline{n}}}^{a}.\end{split}\eeq The field strength
$R_{\underline{m}\underline{n}}$ has only one irreducible
component, \YoungAA{10}. The field strength
$R^{a|bc}={R_{\underline{m}\underline{n}}}^ae^{b\underline{m}}
e^{c\underline{n}}$ decomposes into $\YoungAAA{20}\ \oplus\
{\YoungBA{15}}\ \oplus\ {\YoungA{}}$, with
$\YoungAAA{20}=R^{[a|bc]}$, $\YoungBA{15}\oplus\ {\YoungA{}} =
R^{(a|b)c}$ and\ \
 $\YoungA{}=R^{m|\phantom{m}a}_{\phantom{m|}m}$.

{}From (\ref{SpinTwoSGaugedStrengths}) we see that it is possible
to impose the gauge invariant constraint
\begin{equation}\label{SpinTwoConstraint}
R_{\underline{m}\underline{n}}=0\,,
\end{equation}
which gets rid of the auxiliary field  $\omega^{[a|b]}$. Note that
analogous  constraints express auxiliary fields via derivatives of
the other fields in more complicated systems.

The Bianchi identities (\ref{SpinTwoSBianchi}) reduce to
\beq\begin{split}
&D_{\underline{m}}R_{\underline{n}\underline{k}}-
e_{\underline{m} b}{R_{\underline{n}\underline{k}}}^b\equiv0,\\
&D_{\underline{m}}{R_{\underline{n}\underline{k}}}^a+\lambda^2
{e_{\underline{m}}}^aR_{\underline{n}\underline{k}}\equiv0.\end{split}\eeq
{}From (\ref{SpinTwoConstraint}) it follows
\beq\begin{split}\label{BianchiSpinTwo} &
e_{\underline{m} b}{R_{\underline{n}\underline{k}}}^b\equiv0,\\
&D_{\underline{m}}{R_{\underline{n}\underline{k}}}^a\equiv0.\end{split}\eeq
The first identity implies that the totally antisymmetric
component  \YoungAAA{10} of the field strength
${R_{\underline{n}\underline{k}}}^a$ is zero. The nonzero part of
the field strength $R_{\{2\}}^A$ is therefore contained in
$C^{[ab],c}\equiv R^{c|[ab]}=D^{[a} h^{b]c}$ which is a
generalized Weyl tensor. Its trace part is
$C^{am,\phantom{m}}_{\phantom{am,}m}=D\cdot h-D\trace{h}$.

Since the gauge invariant tensor $C^{[ab],c}$ contains the first
derivatives of the physical field $h^{aa}$, it is possible to
impose the following second order equations of motion  with the
same structure of
 indices as that of
$h^{aa}$ \beq\label{SpinTwoSEq} \alpha D_m C^{(am,a)}+\beta D^{(a}
C^{a)m,\phantom{m}}_{\phantom{a)m,}m} +\gamma \eta^{aa}D_n
C^{nm,\phantom{m}}_{\phantom{nm,}m}=0\,. \eeq This equation is
equivalent to \beq \alpha (2\square h-DD\cdot
h-{2\lambda^2}(dh-\eta \trace{h}))+\beta(2D^2\trace{h}-DD\cdot
h)+\gamma\eta(D^2\cdot h-\square\trace h)=0. \eeq
 By construction, it is
gauge invariant under (\ref{SpinTwoGaugeLaw}) for arbitrary
$\alpha$, $\beta$ and $\gamma$. The condition that the flat limit
 should give the massless field equations in
Minkowski space, which is equivalent to the requirement that the
massless field gauge symmetry  restores in the flat limit, fixes
the coefficients up to an overall factor $\alpha=a/2$,
$\beta=a/2$,  $\gamma=a$. The resulting equations follow from the
lagrangian \beq L=
C^{[mn],a}C_{[mn],a}-\frac12C^{[ms],}_{\phantom{[ms],}s}
C_{[ms],}^{\phantom{[ms],}s}\,, \eeq equivalent to  the lagrangian
(\ref{ActionWeyl}) at $s=2$. The explicit expression in terms of
$h^{aa}$ is \beq L= \frac12(Dh)^2-(D\cdot
h)^2-\frac12(D\trace{h})^2-\trace{h}D^2\cdot h+\frac{d\lambda^2}2
h^2-\frac{\lambda^2}2 {\trace{h}}^2\,. \eeq

Thus, the $s=2$, $t=0$ model of Section \ref{PMreview} is
demonstrated to describe properly the partially massless spin-2
theory of Deser, Waldron \cite{DeserWaldron} and Zinoviev
\cite{Zinoviev}. Note that, in agreement with the general analysis
of Section \ref{ResultsReview}, because of the lack of indices
carried by the field strength 2-form in the degenerate case of
$t=0$, the lagrangian of this model is formulated in terms of the
tensor $C^{[mn],a}$ rather than in terms of the exterior product
of the field strength 2-forms as for $t>0$ systems.

\subsection{Spin-3} \label{SpinThreeSample}

The case of spin-3 admits partially massless fields either with
t=0 or with $t=1$. The case of $t=0$ is analogous to that
considered in Section \ref{SpinTwoSample} for spin-2 and will not
be discussed here. The more interesting case of $s=3$, $t=1$
partially massless field reveals main features of a general
partially massless theory.

The gauge field of this model takes values in the irreducible
$(A)dS$  tensor representation associated with the Young tableau
\YoungBoldBA{10}. In the manifestly antisymmetric basis it
satisfies the conditions
 $W^{AB,C}=-W^{BA,C}$, $W^{[AB,C]}\equiv0$
and $W^{AM,N}\eta_{MN}\equiv0$ (square brackets denote
antisymmetrization). The gauge transformation law, field strength
and Bianchi identities read
 \beq\begin{split}
&\delta W_{\{1\}}^{[AB],C}=D_0\xi_{\{0\}}^{[AB],C}, \\
&R_{\{2\}}^{[AB],C}=D_0W_{\{1\}}^{[AB],C},\\ &D_0
R_{\{2\}}^{[AB],C}\equiv0.
\end{split} \eeq
The \AdS\ Young tableau \YoungBoldBA{10} decomposes into the
following irreps of \grLorenz\ \beq \YoungBoldBA{15}=\YoungBA{15}\
\oplus\ \YoungB{}\ \oplus\ \YoungAA{15}\ \oplus\ \YoungA{}\,.\eeq
Let us introduce the following notation for the gauge parameters
\bee
\begin{split}
&\xi^a=\xi^{a\bullet,\bullet}_{\{0\}}, \\
&\xi^{ab}=\xi^{[a\bullet,b]}_{\{0\}},\\
&\xi^{a,c}=\xi^{(a\bullet,c)}_{\{0\}}, \\
&\xi^{ab,c}=\xi^{ab,c}_{\{0\}}+\frac{1}{d-1}(\xi^a
\eta^{bc}-\xi^b\eta^{ac}).\end{split}\eee The fields
${\omega_{\n{m}}}^{a}$, ${\omega_{\n{m}}}^{ab}$,
${\omega_{\n{m}}}^{a,c}$, ${\omega_{\n{m}}}^{ab,c}$ and the
corresponding field strengths are defined analogously. In these
terms the gauge transformation laws read \beq\begin{split}
\label{SpinThreeGaugeLaw}\delta
{\omega_{\n{m}}}^a&=D_{\n{m}}\xi^a+3\lambda e_{\n{m} m}\xi^{am}+
\lambda e_{\n{m} m}\xi^{a,m},\\
\delta {\omega_{\n{m}}}^{ab}&=D_{\n{m}} \xi^{ab}+ \frac{\lambda
d}{d-1}{e_{\n{m}}}^{[a}\xi^{b]}+
\frac12\lambda e_{\n{m} m}\xi^{ab,m},\\
\delta {\omega_{\n{m}}}^{a,b}&=D_{\n{m}}\xi^{a,b}-\frac{\lambda d
}{d-1}{e_{\n{m}}}^{(a} \xi^{b)}+\frac{\lambda }{d-1}e_{\n{m}
m}\xi^m\eta^{ab}+
\lambda e_{\n{m} m}\xi^{(am,b)},\\
\delta{\omega_{\n{m}}}^{ab,c}&=D_{\n{m}}\xi^{ab,c}+ \lambda
{e_{\n{m}}}^a\xi^{bc}+\lambda {e_{\n{m}}}^a\xi^{b,c}- \lambda
{e_{\n{m}}}^b\xi^{ac}-\lambda {e_{\n{m}}}^b\xi^{a,c}-
2\lambda {e_{\n{m}}}^c\xi^{ab}+\\
&+\frac{\lambda }{d-1}\eta^{bc}e_{\n{m}
m}(3\xi^{am}+\xi^{a,m})-\frac{\lambda }{d-1}\eta^{ac}\lambda
e_{\n{m} m}(3\xi^{bm}+\xi^{b,m}).\end{split} \eeq  The
decomposition of the fields into  irreps of the Lorentz algebra
\grLorenz, that acts diagonally on the fiber indices $a,b,c\ldots$
and the base indices $\underline{m},\n{n}\ldots$, is as follows
\beq
\begin{cases} {\omega_{\n{m}}}^{a}=\YoungB{}\ \oplus\
\YoungAA{15}\ \oplus \bullet, \\
\YoungB{}\oplus\bullet=\omega^{(a|b)},\
\YoungAA{15}=\omega^{[a|b]},\
\chi\equiv\bullet=\omega^{m|}_{\phantom{m|}m}, \end{cases} \eeq

\beq\begin{cases} {\omega_{\n{m}}}^{a,c}={\YoungBA{15}}_1\ \oplus\
\YoungC{}\ \oplus\ \YoungA{},
\\{\YoungBA{15}}_1=\traceless{\omega^{a,b|c}-\omega^{c,(a|b)}},\
\YoungC{}=\traceless{{\phi}^{abc}\equiv\omega^{(a,b|c)}},\
{\psi_1}^a\frac32\equiv{\YoungA{}}^1=\omega^{a,m|}_{\phantom{a,m|}m},\quad
\trace{\phi}=\psi_1,\end{cases}\eeq

\beq\begin{cases} {\omega_{\n{m}}}^{ab}={\YoungBA{15}}_2\ \oplus
\YoungAAA{20}\ \oplus\ {\YoungA{}}^2,\\
{\YoungBA{15}}_2=\traceless{\omega^{ab|c}-\omega^{[ab|c]}},\
\YoungAAA{20}=\omega^{[ab|c]},\
{\psi_2}^a\frac12\equiv{\YoungA{}}^2=\omega^{am|}_{\phantom{am|}m}
,\end{cases}\eeq

\beq\begin{cases} {\omega_{\n{m}}}^{ab,c}=\YoungCA{15}\ \oplus\
\YoungBAA{15}\ \oplus\
\YoungBB{15}\ \oplus\ \YoungB{}\ \oplus \YoungAA{15}\ ,\\
\YoungB{}=\omega^{(am,c)|}_{\phantom{(am,c)|}m},\
\YoungAA{15}=\omega^{[am,c]|}_{\phantom{[am,c]|}m}.\end{cases}\eeq
Explicit expressions are given here for those components, which
appear in the subsequent analysis; the labels $1,2$ distinguish
between different components of the same symmetry type.

{}From (\ref{SpinThreeGaugeLaw}) it follows that the components
\YoungB{} and \YoungAA{10} of ${\omega_{\n{m}}}^{a}$ and
${\YoungBA{15}}_1$ of ${\omega_{\n{m}}}^{a,c}$ can be gauge fixed
to zero by the Stueckelberg gauge parameters $\xi^{a,c}$,
$\xi^{ab}$ and $\xi^{ab,c}$. We therefore set
  \beq\label{SpinThreeSGauge}
\omega^{[a|b]}=0,\ \qquad \traceless{\omega^{(a|b)}}=0,\ \qquad
\traceless{(\omega^{a,b|c}-\omega^{c,(a|b)})}=0.\eeq  This gauge
is not complete. In the leftover gauge transformations, that leave
invariant the gauge conditions, the parameters $\xi^{a,c}$,
$\xi^{ab}$ and $\xi^{ab,c}$ are expressed in terms of derivatives
of $\xi^a$, which is the only non-Stueckelberg gauge parameter.
The gauge transformation law of the component $\phi^{aaa}$ has the
form (\ref{SpinThreeDepthOne}) of the spin-3 partially massless
theory, i.e., $\phi^{aaa}$ identifies with the dynamical field
$\phi$ of Deser and Waldron.

The field strengths have the form \beq\begin{split}
{R_{\n{m}\n{n}}}^a&=D_{\n{m}}{\omega_{\n{n}}}^a+
3\lambda e_{\n{m} m}{\omega_{\n{n}}}^{am}+\lambda e_{\n{m} m}{\omega_{\n{n}}}^{a,m},\\
{R_{\n{m}\n{n}}}^{ab}&=D_{\n{m}} {\omega_{\n{n}}}^{ab}+
\frac{d\lambda }{d-1}{e_{\n{m}}}^{[a}{\omega_{\n{n}}}^{b]}+
\frac12\lambda e_{\n{m} m}{\omega_{\n{n}}}^{ab,m},\\
{R_{\n{m}\n{n}}}^{a,b}&=D_{\n{m}}{\omega_{\n{n}}}^{a,b}-\frac{d\lambda
}{d-1}{e_{\n{m}}}^{(a} {\omega_{\n{n}}}^{b)}+\frac{\lambda }{d-1}
e_{\n{m} m}{\omega_{\n{n}}}^m\eta^{ab}+
\lambda e_{\n{m} m}{\omega_{\n{n}}}^{(am,b)},\\
{R_{\n{m}\n{n}}}^{ab,c}&=D_{\n{m}}{\omega_{\n{n}}}^{ab,c}+ \lambda
{e_{\n{m}}}^a{\omega_{\n{n}}}^{bc}+ \lambda
{e_{\n{m}}}^a{\omega_{\n{n}}}^{b,c}- \lambda
{{e_{\n{m}}}^b}{\omega_{\n{n}}}^{ac}- \lambda
{e_{\n{m}}}^b{\omega_{\n{n}}}^{a,c}-
2\lambda {e_{\n{m}}}^c{\omega_{\n{n}}}^{ab}+\\
&+\frac{\lambda }{d-1}\eta^{bc}e_{\n{m}
m}(3{\omega_{\n{n}}}^{am}+{\omega_{\n{n}}}^{a,m})-\frac{\lambda
}{d-1}\eta^{ac}e_{\n{m}
m}(3{\omega_{\n{n}}}^{bm}+{\omega_{\n{n}}}^{b,m}).\end{split}\eeq
Let us consider which gauge invariant conditions can be imposed on
the fields by setting to zero some of components of the field
strengths. These may include either constraints, that express
auxiliary fields in terms of derivatives of the dynamical fields,
or impose differential field equations on the dynamical fields.

The simplest constraint is ${R_{\n{m}\n{n}}}^a=0$. Taking into
account the gauge conditions (\ref{SpinThreeSGauge}), it implies
\beq\label{firstof}\YoungAAA{20}=\omega^{[ab|c]}=0,\eeq
\beq\label{ZeroHook}
{\YoungBA{15}}_2=\traceless{\omega^{ab|c}-\omega^{[ab|c]}}=0\,,\eeq
\beq\label{constr0} \frac{d-1}d D^a\chi-\frac32 {\psi_2}^a+\frac32
{\psi_1}^a=0.\eeq The constraint (\ref{constr0}) can be used to
express the auxiliary field ${\psi_2}^a$ in terms of ${\psi_1}^a$
and the first derivative of the scalar field $\chi$. The
constraint (\ref{ZeroHook}) in the gauge (\ref{SpinThreeSGauge})
implies that the hook-symmetry parts of $\omega^{ab}$ and
$\omega^{a,b}$ are zero. As a result one is left with traces
$\chi$, ${\psi_2}^a$ of $\omega^a$, $\omega^{ab}$, with the
totally symmetric component $\phi^{aaa}$ of $\omega^{a,b}$, with
its trace ${\psi_1}^a$ and the field $\omega^{ab,c}$, which has
not been yet considered.

Let us now impose the condition ${R_{\n{m}\n{n}}}^{a,b}=0$. It
implies that
\beq\YoungBAA{15}=\traceless{R^{a,[b|bb]}}=\traceless{\omega^{[bb,a|b]}}=0,\eeq
\beq\label{constr31}
\omega^{(am,a|a)}=-D^m\phi^{aaa}+D^{(a}\omega^{a,a)|m}, \eeq
expressing the auxiliary fields
$\omega^{(am,a|a)}=\YoungCA{15}\oplus\ \YoungB{}
\oplus\YoungAA{15}$ in terms of the first derivatives of
$\phi^{aaa}$ and setting the component
$\YoungBAA{15}=\omega^{[aa,b|a]}$ to zero.

The condition that the projection of $R^{[ab],c}$ to the
components of the symmetry \YoungC{} and \YoungA{} is zero (i.e.,
$R^{m(a,a|a)}{}_m=0$ and $R^{st,a|}{}_{st}=0$) gives the following
equation
 \beq\label{PhiEquation} \square\phi-DD\cdot
\phi+\frac12DD\psi_1-3\lambda^2\phi+\eta\lambda^2\psi_1-
\eta\lambda^2\frac{d-2}{3d}D\chi=0\,, \eeq
 which is the dynamical equation on $\phi$ and its trace $\psi_1$.

The condition ${R_{\n{m}\n{n}}}^{ab}=0$ in the gauge
(\ref{SpinThreeSGauge}) sets to zero the component $\YoungBB{15}$\
\ of $\omega^{ab,c}$ and
$\traceless{R^{(am|a)}_{\phantom{(am|a)}m}}=0$ implies  \beq
\label{constrDW} \traceless{D^{(a}{\psi_1}^{a)}-D_m\phi^{maa}+
\frac{d-2}{3d\lambda}D^aD^a\chi}=0,\eeq while ${R_{st}}^{st}=0$
imposes the equation \beq D_m {\psi_2}^m+\lambda d\chi=0\,. \eeq
Using  (\ref{constr0}) to exclude ${\psi_2}^a$, we obtain
\beq\label{ChiEquation} \frac{2(d-1)}{3d}\square\chi+\lambda D_m
{\psi_1}^m+d\lambda^2\chi=0\,, \eeq which is the equation on
$\chi$.

As a result, one has the dynamical equations (\ref{PhiEquation}),
(\ref{ChiEquation}) on $\phi^{aaa}$, its trace ${\psi_1}^a$ and
$\chi$, plus one differential constraint (\ref{constrDW}), which
corresponds to equation (\ref{SpinThree2}) of the approach of
Deser and Waldron.

Consider a \AdS-covariant, gauge invariant action of the form \beq
S=\frac1{2\lambda}\int_{M^d} \left(a_1 R_{\{2\}}^{[A_1
A_2],B}R_{\{2\}}^{[A_3 A_4],}{}_B +
a_2R_{\{2\}}^{[A_1A_2],C}R_{\{2\}}^{[A_3A_4],C}V_{C(2)}\right)\epsilon_{A_1
\ldots A_{d+1}}\ V^{A_{5}} E^{A_{6}}\ldots E^{A_{d+1}}\,. \eeq In
terms of Lorentz components rescaled according to (\ref{Rescaling}),
its variation  turns out to be \beq
\begin{split}\delta S=\int_{M^d} &\alpha ( \delta
\omega^{[a_1 a_2]} R^{a_3}+\delta\omega^{a_1} R^{[a_2 a_3]} )+\\
+&\beta (\delta\omega^{[a_1 a_2],\ b}{R^{a_3,}}_{
b}+\delta\omega^{a_1,\ b}{R^{[a_2 a_3],}}_{ b})+\\
+&\frac1{\lambda}\gamma (\delta\omega^{[a_1 a_2],\ b}{R^{a_3}}_{
b}+\delta\omega^{[a_1 b]}{R^{[a_2 a_3],}}_{b}) \varepsilon_{a_1
\ldots a_d} h^{a_4} \ldots h^{a_d},\end{split}\eeq where
 \beq
\alpha=\frac{2d}{d-1}( a_1+a_2),\ \ \beta= a_1,\ \ \gamma= a_1
-\frac2{d-3} a_2\,.\eeq To have correct flat limit we set
$\gamma=0$. As a result $a_2=\frac{d-3}2 a_1$, which is in
accordance with the general formula (\ref{Coefficientsb}). The
resulting action gives the following equations of motion \beq
\frac{\delta S}{\delta {\omega_{\n{m}}}^{[ab]}}=0
\Longleftrightarrow {R_{\n{m}\n{n}}}^a=0 ,\eeq \beq \frac{\delta
S}{\delta \chi}=0 \Longleftrightarrow {R_{st}}^{st}=0,\eeq  \beq
\frac{\delta S}{\delta {\omega_{\n{m}}}^{[ab],c}}=0
\Longleftrightarrow {R_{mn}}^{a,c}=0, \eeq \beq \frac{\delta
S}{\delta \phi^{aaa}}=0 \Longleftrightarrow
{R}^{(am,a|a)}_{\phantom{(am,a|a)}m}=0\,. \eeq These equations are
equivalent to the conditions (\ref{firstof})-(\ref{ChiEquation})
on the field strengths, which have been just shown to describe
properly the spin-3 partially massless theory of depth one.

\section{Discussion}\setcounter{equation}{0}  \label{Conclusion}

In this paper, partially massless fields in $(A)dS$ space studied
in \cite{FirstPMWorks}-\cite{Dolan:2001ih} are shown to admit a
natural formulation in terms of gauge 1-forms that take values in
irreducible tensor representations of the  algebra
$(o(d,1))o(d-1,2)$, described by the various two-row Young
tableaux. The case of rectangular two-row Young tableaux
corresponds to the  massless fields \cite{Vasiliev:2001wa}.
Different non-rectangular Young tableaux correspond to different
partially massless fields. The manifestly gauge invariant free
lagrangians are bilinears of gauge invariant field strengths.

One of the applications of the obtained results is that they
impose strong restrictions on possible HS algebras. Actually,
since partially massless fields in $AdS_d$ correspond to
non-unitary representations of $o(d-1,2)$, for a HS theory in
$AdS_d$ to be unitary, it should be free of gauge fields
associated with  partially massless fields. This means that a HS
algebra $g$ considered as a $o(d-1,2)$-module with respect to its
adjoint action in $g$ should not contain submodules described by
the corresponding Young tableaux. This condition is indeed
satisfied in the case of HS algebras underlying HS theories of
totally symmetric fields \cite{Vasiliev:2004cm}.
 The simplest example of a HS algebra,
which is associated with the supersymmetric extension of the
theory of symmetric HS fields studied in \cite{Vasiliev:2004cm},
also satisfies this condition.  It should be true, however, for
any other unitary HS theory including those that contain generic
mixed symmetry fields  \cite{Alkalaev:2005kw},
\cite{Curtright:1980yj}, \cite{ZinovievMSPM}, thus imposing
nontrivial restrictions on a structure of a HS algebra and,
therefore, on the spectra of HS gauge fields that may appear in
consistent HS gauge theories.

It would be interesting to extend the results of this paper to the
case of general mixed symmetry partially massless bosonic and
fermionic  fields, originally discussed in \cite{ZinovievMSPM}.
Taking into account the results of \cite{Alkalaev:2003qv}, where
it has been shown that massless fields of a general symmetry type
are described by $p$-form gauge fields that carry a representation
of \grAdS\ described by a Young tableau with at least $p+1$ cells
in the shortest column, it is natural to conjecture that partially
massless fields as well as the ``nonunitary  massless fields" with
the gauge parameters ruled out in \cite{Metsaev:1997nj} are
described in terms of $p$-form gauge fields that are
\grAdS-modules depicted by Young tableaux with at most $p$ cells
in the shortest column. The analysis of this problem, that we
leave for a future publication, will provide the full list of
unitary and non-unitary free HS models described by gauge $p$-form
fields that take values in various finite-dimensional
\grAdS-modules.

\section*{Acknowledgements}
The authors acknowledge with gratitude the collaboration of
V.Didenko at the early stage of this work. E.S. is grateful to
K.B.Alkalaev for useful discussions. The work was supported in
part by grants RFBR No. 05-02-17654, LSS No. 1578.2003-2 and INTAS
No. 03-51-6346. The work of E.S. was supported in part by the
scholarship of Dynasty Foundation.

\newpage
\appendix

\section{Appendix A: Notation}\setcounter{equation}{0} \label{Notation}
\renewcommand{\theequation}{\Alph{section}.\arabic{equation}}
\setcounter{section}{1}
\subsection{Index conventions}
Latin indices $A,B,... =  0,1,...,d$ are
  vector indices of the isometry algebra
\grAdS\  or $o(d,1)$. Indices $a,b,...= 0\,,\ldots, d-1$ are
vector indices of the $d$-dimensional Lorentz algebra \grLorenz.
For definiteness, we consider the \AdS\ case of \grAdS. For the
decomposition of irreps of \grAdS\ into irreps of \grLorenz\ we
use notations $A=\{a,\bullet\}$, where
 $\bullet$ denotes
the extra value of the \grAdS\ vector index compared to the
  \grLorenz\ vector index.

Underlined Latin indices
$\underline{m},\underline{n},\underline{k},...= 0,1,\ldots d-1$
are the indices of differential forms in the space-time base
manifold. $d=dx^{\underline{m}}\pl_{\underline{m}}$ is the de Rham
differential. To simplify  formulae we often omit the exterior
differentials $dx^{\underline{m}}$ and the wedge product symbol
$\wedge$. The rank of a differential form is indicated as a braced
subscript. For example, $\omega_{\{r\}}$ denotes a $r$-form
$\omega_{\underline{m}_1\ldots\underline{m}_r}$.

$\eta_{AB}$ and $\eta_{ab}$  are, respectively, \grAdS\ and
\grLorenz invariant metrics. We use the mostly minus signature,
i.e., $\eta_{AB}=\mbox{diag}(\overbrace{+-...-+}^{d+1})$ and
  $\eta_{ab}=\mbox{diag}(\overbrace{+-...-}^{d})$.
$\epsilon_{A_1\ldots A_{d+1}}$ and $\epsilon_{a_1\ldots a_{d}}$
are Levi-Civita tensors of the algebras \grAdS\ and $o(d-1,1)$,
respectively. ${e_{\underline{m}}}^a$ is the vielbein 1-form.
 The metric tensor
$g_{\underline{m}\underline{n}}={e_{\underline{m}}}^a$
${e_{\underline{n}}}^b \eta_{ab}$ is required to be nondegenerate.
${\lambda}^{-1}$  is the radius of  $(A)dS_d$.

To simplify formulae we use multi-index notation. Namely, a group
of indices of any kind (in particular, vector indices of the
algebras \grAdS\ and \grLorenz) in which some tensor $C^{a_1 a_2
\ldots a_n, b,c\ldots}$ is symmetric, that is \beq C^{a_1 a_2
\ldots a_i \ldots a_j \ldots a_n, b,c,\ldots}=C^{a_1 a_2 \ldots
a_j \ldots a_i \ldots a_n, b,c,\ldots}\eeq is denoted $C^{a(n),
b,c,\ldots}\equiv C^{(a_1 a_2 \ldots a_n), b,c,\ldots}$. The
similar notation
 $C^{a[n],...,}\equiv C^{[aa...a],...,}$ is used for antisymmetric indices.
More generally, symmetrized upper or lower indices are denoted by
the same letter. We use the normalization in which the
symmetrization operator $\mathbf{ \widehat{S}}$ is a projector
$\mathbf{ \widehat{S}^2} = \mathbf {\widehat{S}}$. For example,
for any vector $V^a$ and a totally symmetric rank-$n$ tensor
$C^{a(n)}$ \beq V^a C^{a(n)}=\frac1{n+1}\sum_{i=1}^{i=n+1} V^{a_i}
C^{a_1 \ldots \widehat{a_i} \ldots a_{n+1}}, \eeq where the hatted
index has to be omitted.

\subsection{Condensed Notation} The traceless part of a tensor $T$ is
denoted $\traceless{T}$. For example, for a rank-2 tensor,
$\phi_{ab}$ \beq
\traceless{\phi_{ab}}=\phi_{ab}-\frac1d\eta_{ab}\phi^c_{\phantom{c}c}.\eeq
A tensor field that has the symmetry properties depicted by the
Young tableau composed of rows of lengths $n_1,\ldots,n_p$ is
denoted $\phi^{\{n_1,\ldots,n_p\}}$.

The following notation is often used for a totally symmetric
rank-$s$ tensor field $\phi_{\underline{m}_1 \ldots
\underline{m}_s}(x)$ and its derivatives  \beq \phi^{\{s\}} \equiv
\phi_{\underline{m}_1 \ldots \underline{m}_s}(x),\eeq \beq
\pl\phi^{\{s\}} \equiv
s\pl_{(\underline{m}_1}\phi_{\underline{m}_2
\ldots\underline{m}_{s+1})}(x), \eeq \beq
\Box\phi^{\{s\}}\equiv\pl^{\underline{m}}\pl_{\underline{m}}\phi^{\{s\}},\eeq
  \beq \pl\cdot\phi^{\{s\}} \equiv
\pl^{\underline{m}}{\phi}_{\underline{m} \underline{m}_2 \ldots
\underline{m}_s}(x),\eeq \beq \pl^2\cdot\phi^{\{s\}}\equiv
\pl\cdot(\pl\cdot\phi^{\{s\}}) ,\eeq \beq
\trace{{\phi^{\{s\}}}}\equiv
\phi^{\underline{m}}_{\phantom{\underline{m}}\underline{m}
\underline{m}_3 \ldots \underline{m}_s}(x),\eeq \beq
({\pl\cdot{\phi^{\{s\}}}})^2\equiv\pl^{\underline{n}}
{\phi}_{\underline{n} \underline{m}_2 \ldots \underline{m}_s}(x)
\pl_{\underline{k}} {\phi}^{\underline{k} \underline{m}_2 \ldots
\underline{m}_s}(x), \eeq  \beq \eta\phi^{\{s\}}\equiv
\frac{(s+2)(s+1)}{2}\eta_{(\underline{m}_1\underline{m}_2}\phi_{\underline{m}_3
\ldots \underline{m}_{s+2})}.\eeq

\section{Appendix B: Young tableaux}  \setcounter{equation}{0}\label{AuxInfo}
\setcounter{section}{2} A Young tableau $\{ n_i\}$ ($i=1,2,...,p$)
is a set of positive integers such that $n_i\geq n_j >0, \ i>j$.
It can be depicted as a set of rows of length $n_i$.

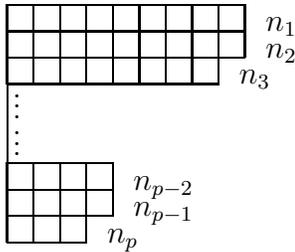
\YoungGeneralized \\ Each box corresponds to an index of a tensor

\beq C^{A_1 A_2 \ldots A_{n_1},B_1 B_2 \ldots
B_{n_2},\ldots\ldots,Z_1 Z_2 \ldots Z_{n_p}}\,, \eeq that has the
following properties
\begin{enumerate}
     \item The tensor $C^{A_1 A_2 \ldots A_{n_1},B_1 B_2 \ldots
     B_{n_2},\ldots\ldots,Z_1 Z_2 \ldots Z_{n_p}}$
is symmetric in indices $A_1\ldots A_{n_1}$, $B_1 \ldots B_{n_2}$
and so forth,
     that is, in multi-index notation, it is $C^{A(n_1),B(n_2),\ldots,Z(n_p)}$.

     \item Symmetrization of all indices of any group with an
     index of a subsequent groups is identically zero
     \beq C^{A(n_1),\ldots,(F(n_k),\ldots,F)N(n_m-1),\ldots,Z(n_p)}=0.\eeq

\end{enumerate}

A scalar is symbolized by $\bullet$.

Different symmetry types of tensors are characterized by  Young
tableaux. One can use different bases in the vector space of
tensors of a given type. The one we  mostly use in this paper is
that with explicit symmetrizations. Alternatively,
 one can use the antisymmetric basis, which is used in particular
in Section \ref{SpinThreeSample} of this paper.  (For more detail,
see e.g. \cite{Bekaert:2005vh}.)

Since the metric and the Levi-Civita tensor $\epsilon_{A_1\ldots A_d
}$ are invariant tensors of the algebra $o(r,q)$, to single out its
irreps the tracelessness condition, the condition that the height of
any column does not exceed $[d/2]$ have to be imposed. For even $d$
and appropriate signature the (anti)selfduality condition has to be
imposed on the tensor representations associated with Young tableaux
that have at least one column of height $d/2$.

The decomposition of the tensor product of any irrep of $sl_d$
with its vector representation admits a natural description in
terms of Young tableaux. Namely, it contains all tableaux that can
be obtained by adding a box to a given Young tableau. For example,

\beq \label{TensorProduct} \YoungBA{15}\ \otimes\ \YoungA{}
=\YoungCA{15}\oplus\YoungBB{15}\oplus\YoungBAA{15}.\eeq

In the case of  $o(r,q)$, Young tableaux are associated with
traceless tensors of one or another symmetry type. The tensor
product decomposition results from either adding or cutting one
box in various ways. The cutted  diagrams result from traces
between the tensor product factors. For example, the $o(r,q)$
decomposition analogous to (\ref{TensorProduct}) is \beq
\YoungBA{15}\ \otimes\
\YoungA{}=\YoungCA{15}\oplus\YoungBB{15}\oplus\YoungBAA{15}\oplus\
\YoungB{}\oplus\YoungAA{10}\,. \eeq

To decompose an irrep of \grAdS\ into irreps of its subalgebra
\grLorenz\ it is convenient to identify the latter with the
stability subalgebra of some nonzero time-like vector $V^A$. The
mnemonic is that  to obtain a set of tableaux of \grLorenz\ one
has to cut off boxes of the Young tableau of \grAdS\ in all
possible ways so that no two boxes were cut off in the same
column.  For example, for the \grAdS\ tableau \YoungCA{10} we
obtain the following set of \grLorenz\ tableaux \beq \YoungCA{10}\
\oplus\ \YoungBA{10}\ \oplus\ \YoungC{}\oplus\ \YoungAA{10}\
\oplus\ \YoungB{}\oplus\ \YoungA{}.\eeq

In this paper, fields, gauge parameters and field strengths are
$p$-forms with fiber indices in finite-dimensional modules of
\grLorenz\ or \grAdS\,  described by various Young tableaux.
Replacing the form index of a 1-form by the fiber one with the
help of the vielbein ${e_{\underline{m}}}^{a}$ (e.g.,
$\omega^{a(3),b|c}={\omega_{\underline{m}}}^{a(3),b}e^{\underline{m}|c}$),
the resulting fiber tensor decomposes into irreps of \grLorenz\
 according to the general rule for the tensor product
with the vector representation. For instance, one obtains for
$\omega^{a(3),b|c}$ \beq \YoungCA{15}\ \otimes\
\YoungA{}=\YoungDA{15}\ \oplus\YoungCB{15}\
\oplus\YoungCAA{20}\oplus\YoungBA{15}\oplus\ \YoungC{}.\eeq In
particular, here  \YoungC{} corresponds to
${\omega_{\underline{m}}}^{a(3),b} {e^{\underline{m}}}_b$.

\section{Appendix C: Coefficients}
\setcounter{equation}{0}\label{Coefficients}
\setcounter{section}{3} The final result is \beq
g(k,m)=\sqrt{h(k,m) \frac{k+1}{k(d+k-2)}}\ ,\eeq \beq
G(k,m)=\sqrt{H(k,m) \frac{k-m}{(d+m-3)(k-m+1)}}\ ,\eeq \beq
f(k,m)=\sqrt{h(k-1,m) \frac{(k-1)(d+k-3)}{k}}\ ,\eeq \beq
F(k,m)=\sqrt{H(k,m-1) \frac{(d+m-4)(k-m+2)}{k-m+1}}\ ,\eeq where
\bee \label{h}
h(k,m)=\frac{(k-t+1)(d+k+t-2)(s-k-1)(d+s+k-2)}{(k-m+1)(d+k+m-2)(d+2k)},
\\ H(k,m)=\frac{(s-m)(d+s+m-3)(t-m)(d+t+m-3)}{(k-m+1)(d+u+m-2)(d+2m-2)}.\eee

These coefficients satisfy the following relations
\bee f(k+1,m)=\frac{(d+k-2)k}{(k+1)}g(k,m), \\
F(k,m+1)=\frac{(d+m-3)(k-m+1)}{(k-m)}G(k,m)\, \eee which insure
that the operators $\sigma^{1,2}_{+}$ are conjugated to
$\sigma^{1,2}_{-}$ with respect to the scalar product
(\ref{MyScalarProduct}). Also note that
\bee h(k,m)=f(k+1,m)g(k,m), \\
H(k,m)=F(k,m+1)G(k,m)\,. \eee


\newpage


\begin{thebibliography}{99}

\bibitem{FirstPMWorks}
   S.~Deser and R.~I.~Nepomechie,
   Phys.\ Lett.\ B {\bf 132} (1983) 321;
   Annals Phys.\  {\bf 154} (1984) 396.

\bibitem{Higuchi}
  A.~Higuchi,
YTP-85-22;
  Nucl.\ Phys.\ B {\bf 282}, 397 (1987);
  J.\ Math.\ Phys.\  {\bf 28}, 1553 (1987)
  [Erratum-ibid.\  {\bf 43}, 6385 (2002)];
  Class.\ Quant.\ Grav.\  {\bf 6}, 397 (1989);
  Nucl.\ Phys.\ B {\bf 325}, 745 (1989).


\bibitem{Bengtsson:1994vn}
  I.~Bengtsson,
  J.\ Math.\ Phys.\  {\bf 36}, 5805 (1995)
  [arXiv:gr-qc/9411057].

\bibitem{DeserWaldron}
  S.~Deser and A.~Waldron,
  Phys.\ Rev.\ Lett.\  {\bf 87}, 031601 (2001)
  [arXiv:hep-th/0102166];
   Nucl.\ Phys.\ B {\bf 607} (2001) 577
   [arXiv:hep-th/0103198];
  Phys.\ Lett.\ B {\bf 508}, 347 (2001)
  [arXiv:hep-th/0103255];
  Phys.\ Lett.\ B {\bf 513}, 137 (2001)
  [arXiv:hep-th/0105181];
   Nucl.\ Phys.\ B {\bf 662} (2003) 379
   [arXiv:hep-th/0301068];
   ``Conformal invariance of partially massless higher spins'',
   arXiv:hep-th/0408155.

\bibitem{Zinoviev}
   Y.~M.~Zinoviev,
   ``On massive high spin particles in (A)dS'',
   arXiv:hep-th/0108192.



\bibitem{Buchbinder:2000fy}
  I.~L.~Buchbinder, D.~M.~Gitman and V.~D.~Pershin,
  Phys.\ Lett.\ B {\bf 492}, 161 (2000)
  [arXiv:hep-th/0006144].

\bibitem{Hallowell:2005np}
  K.~Hallowell and A.~Waldron,
  Nucl.\ Phys.\ B {\bf 724}, 453 (2005)
  [arXiv:hep-th/0505255].


\bibitem{Dolan:2001ih}
  L.~Dolan, C.~R.~Nappi and E.~Witten,
  JHEP {\bf 0110}, 016 (2001)
  [arXiv:hep-th/0109096].


\bibitem{Evans}
 N.~T.~Evans,
J.\ Math.\ Phys.\ {\bf 8},  170 (1967).

\bibitem{Metsaev:1997nj}
R.R. Metsaev, Phys.\ Lett. B \textbf{354} (1995) 78;
  ``Arbitrary spin massless bosonic fields in d-dimensional anti-de Sitter
  space,''
  arXiv:hep-th/9810231.

\bibitem{Angelopoulos:1997ij}
  E.~Angelopoulos and M.~Laoues,
  Rev.\ Math.\ Phys.\  {\bf 10}, 271 (1998)
  [arXiv:hep-th/9806100].

\bibitem{Brink:2000ag}
   L.~Brink, R.~R.~Metsaev and M.~A.~Vasiliev,
   Nucl.\ Phys.\ B {\bf 586}, 183 (2000)
   [arXiv:hep-th/0005136].

\bibitem{Ferrara:2000nu}
  S.~Ferrara and C.~Fronsdal,
  ``Conformal fields in higher dimensions,''
  arXiv:hep-th/0006009.

\bibitem{Vasiliev:2004cm}
  M.~A.~Vasiliev,
  JHEP {\bf 0412}, 046 (2004)
  [arXiv:hep-th/0404124].


\bibitem{Dolan:2005wy}
  F.~A.~Dolan,
  ``Character formulae and partition functions in higher dimensional conformal
  field theory,''
  arXiv:hep-th/0508031.


\bibitem{Vasiliev:1980as}
   M.~A.~Vasiliev,
   Yad.\ Fiz.\  {\bf 32}, 855 (1980).

\bibitem{Vasiliev:1986td}
   M.~A.~Vasiliev,
   Fortsch.\ Phys.\  {\bf 35}, 741 (1987)
   [Yad.\ Fiz.\  {\bf 45}, 1784 (1987)].

\bibitem{Lopatin:1987hz}
   V.~E.~Lopatin and M.~A.~Vasiliev,
   Mod.\ Phys.\ Lett.\ A {\bf 3}, 257 (1988).


\bibitem{Vasiliev:2001wa}
   M.~A.~Vasiliev,
   Nucl.\ Phys.\ B {\bf 616} (2001) 106
   [Erratum-ibid.\ B {\bf 652} (2003) 407]
   [arXiv:hep-th/0106200].

\bibitem{Alkalaev:2003qv}
   K.~B.~Alkalaev, O.~V.~Shaynkman and M.~A.~Vasiliev,
   Nucl.\ Phys.\ B {\bf 692} (2004) 363
   [arXiv:hep-th/0311164].

\bibitem{Dirac:1936} Dirac, P.~A.~M.,
Royal Society of London Proceedings Series A, 155, 447, 1936.


\bibitem{Fierz:1939ix}
   M.~Fierz and W.~Pauli,
   Proc.\ Roy.\ Soc.\ Lond.\ A {\bf 173} (1939) 211.

\bibitem{Singh:1974qz}
   L.~P.~S.~Singh and C.~R.~Hagen,
   Phys.\ Rev.\ D {\bf 9} (1974) 898.


\bibitem{Bianchi:2006gk}
  M.~Bianchi and F.~Riccioni,
  ``Massive higher spins and holography,''
  arXiv:hep-th/0601071.


\bibitem{Buchbinder:2005ua}
  I.~L.~Buchbinder and V.~A.~Krykhtin,
  Nucl.\ Phys.\ B {\bf 727}, 537 (2005)
  [arXiv:hep-th/0505092].

\bibitem{Buchbinder:2005cf}
  I.~L.~Buchbinder and V.~A.~Krykhtin,
  ``BRST approach to higher spin field theories,''
  arXiv:hep-th/0511276.



\bibitem{Fronsdal:1978rb}
   C.~Fronsdal,
   Phys.\ Rev.\ D {\bf 18} (1978) 3624.



\bibitem{Buchbinder:2004gp}
  I.~L.~Buchbinder, V.~A.~Krykhtin and A.~Pashnev,
  Nucl.\ Phys.\ B {\bf 711}, 367 (2005)
  [arXiv:hep-th/0410215].

\bibitem{MasslessHSAlternative}
  D.~Francia and A.~Sagnotti,
  Phys.\ Lett.\ B {\bf 543}, 303 (2002)
  [arXiv:hep-th/0207002];
  Class.\ Quant.\ Grav.\  {\bf 20}, S473 (2003)
  [arXiv:hep-th/0212185];
  Phys.\ Lett.\ B {\bf 624}, 93 (2005)
  [arXiv:hep-th/0507144].




\bibitem{MacDowell:1977jt}
   S.~W.~MacDowell and F.~Mansouri,
   Phys.\ Rev.\ Lett.\  {\bf 38}, 739 (1977)
   [Erratum-ibid.\  {\bf 38}, 1376 (1977)].

\bibitem{Bekaert:2005vh}
   X.~Bekaert, S.~Cnockaert, C.~Iazeolla and M.~A.~Vasiliev,
   ``Nonlinear higher spin theories in various dimensions'',
   arXiv:hep-th/0503128.

\bibitem{Stelle:1979aj}
   K.~S.~Stelle and P.~C.~West,
   Phys.\ Rev.\ D {\bf 21} (1980) 1466.

\bibitem{rev}  
  M.~A.~Vasiliev,
  `Higher spin gauge theories: Star-product and AdS space,''
  arXiv:hep-th/9910096;
  Comptes Rendus Physique {\bf 5}, 1101 (2004)
  [arXiv:hep-th/0409260];\\
E.~Sezgin and P.~Sundell,
  Nucl.\ Phys.\ B {\bf 644}, 303 (2002)
  [Erratum-ibid.\ B {\bf 660}, 403 (2003)]
  [arXiv:hep-th/0205131];\\
  D.~Sorokin,
  AIP Conf.\ Proc.\  {\bf 767}, 172 (2005)
  [arXiv:hep-th/0405069];\\
  N.~Bouatta, G.~Compere and A.~Sagnotti,
  ``An introduction to free higher-spin fields,''
  arXiv:hep-th/0409068.

\bibitem{Skvortsov:2006}
  E.~D.~Skvortsov, M.~A.~Vasiliev,
  in preparation.


\bibitem{Alkalaev:2005kw}
  K.~B.~Alkalaev, O.~V.~Shaynkman and M.~A.~Vasiliev,
  JHEP {\bf 0508}, 069 (2005)
  [arXiv:hep-th/0501108].

\bibitem{Curtright:1980yj}
  T.~Curtright and P.~G.~O.~Freund,
  Nucl.\ Phys.\ B {\bf 172}, 413 (1980);

  T.~Curtright,
  Phys.\ Lett.\ B {\bf 165}, 304 (1985);

  C.~S.~Aulakh, I.~G.~Koh and S.~Ouvry,
  Phys.\ Lett.\ B {\bf 173}, 284 (1986);

  J.~M.~F.~Labastida and T.~R.~Morris,
  Phys.\ Lett.\ B {\bf 180}, 101 (1986).
  J.~M.~F.~Labastida,
  Phys.\ Rev.\ Lett.\  {\bf 58}, 531 (1987);
  Nucl.\ Phys.\ B {\bf 322}, 185 (1989);


  M.~Dubois-Violette and M.~Henneaux,
  Lett.\ Math.\ Phys.\  {\bf 49}, 245 (1999)
  [arXiv:math.qa/9907135];
  Commun.\ Math.\ Phys.\  {\bf 226}, 393 (2002)
  [arXiv:math.qa/0110088];

  C.~Burdik, A.~Pashnev and M.~Tsulaia,
  Mod.\ Phys.\ Lett.\ A {\bf 16}, 731 (2001)
  [arXiv:hep-th/0101201];

  P.~de Medeiros and C.~Hull,
  JHEP {\bf 0305}, 019 (2003)
  [arXiv:hep-th/0303036];

  P.~de Medeiros,
  Class.\ Quant.\ Grav.\  {\bf 21}, 2571 (2004)
  [arXiv:hep-th/0311254];

  X.~Bekaert and N.~Boulanger,
  Commun.\ Math.\ Phys.\  {\bf 245}, 27 (2004)
  [arXiv:hep-th/0208058];

  C.~C.~Ciobirca, E.~M.~Cioroianu and S.~O.~Saliu,
  Int.\ J.\ Mod.\ Phys.\ A {\bf 19}, 4579 (2004)
  [arXiv:hep-th/0403017];

  R.~R.~Metsaev,
  Class.\ Quant.\ Grav.\  {\bf 22}, 2777 (2005)
  [arXiv:hep-th/0412311];

  K.~B.~Alkalaev,
  ``Mixed-symmetry gauge fields in AdS(5),''
  arXiv:hep-th/0501105;
  Theor.\ Math.\ Phys.\  {\bf 140}, 1253 (2004)
  [Teor.\ Mat.\ Fiz.\  {\bf 140}, 424 (2004)]
  [arXiv:hep-th/0311212];

\bibitem{ZinovievMSPM}
  Y.~M.~Zinoviev,
  ``On massive mixed symmetry tensor fields in Minkowski space and (A)dS,''
  arXiv:hep-th/0211233;
  ``First order formalism for mixed symmetry tensor fields,''
  arXiv:hep-th/0304067;
  ``First order formalism for massive mixed symmetry tensor fields in
  Minkowski and (A)dS spaces,''
  arXiv:hep-th/0306292.

\end{thebibliography}
\end{document}